\documentclass[a4paper,11pt]{article}
\usepackage{jheppub}
\usepackage[T1]{fontenc} % if needed
\usepackage{epsfig}
\usepackage{bm}% bold math
\usepackage{color}
\usepackage[utf8]{inputenc} % pour accents de clavier 
\usepackage[T1]{fontenc}    

\def\ben{\begin{enumerate}}
\def\een{\end{enumerate}}
\def\bit{\begin{itemize}}
\def\eit{\end{itemize}}
\def\beq{\begin{equation}}
\def\eeq{\end{equation}}
\def\bea{\begin{eqnarray}}
\def\eea{\end{eqnarray}}
\def\bq{\begin{quote}}
\def\eq{\end{quote}}
\def \lsim{\mathrel{\vcenter
     {\hbox{$<$}\nointerlineskip\hbox{$\sim$}}}}
\def \gsim{\mathrel{\vcenter
     {\hbox{$>$}\nointerlineskip\hbox{$\sim$}}}}
\def\gappeq{\mathrel{\rlap {\raise.5ex\hbox{$>$}}
{\lower.5ex\hbox{$\sim$}}}}
\def\lappeq{\mathrel{\rlap{\raise.5ex\hbox{$<$}}
{\lower.5ex\hbox{$\sim$}}}}

\def\LNP{\Lambda_{NP}}

\def\meg{\mu \to e \gamma}
\def\megL{\mu \to e_L \gamma}
\def\megg{\mu \to e \gamma \gamma}
\def\muc{\mu A\to e A}
\def\mucL{\mu A\to e_L A}
\def\mec{\mu \! \to \! e~ {\rm conversion}}
\def\teg{\tau \to e \gamma}
\def\tmg{\tau \to \mu \gamma}

\def\tlg{\tau \to \ell \gamma}

\def\meee{\mu \to e \bar{e} e}
\def\meeeL{\mu \to e_L \overline{e_L} e_L}
\def\tlll{\tau \to 3 \ell }
\def\teee{\tau \to e \bar{e} e }
\def\tmee{\tau \to \mu \bar{e} e }
\def\temm{\tau \to e  \bar{\mu} \mu }
\def\tmmm{\tau \to \mu \bar{\mu} \mu }
\def\teme{\tau \to e \bar{\mu} e }

\def\a{\alpha}
\def\b{\beta}
\def\g{\gamma}

\def\m{\mu}

\def\s{\sigma}

\title{\boldmath  Completeness and
Complementarity for  $\meg$, $\meee$ and $\mu A \to eA$ }

\author{ S Davidson}

\affiliation{
LUPM, CNRS,
Universit\'{e} Montpellier,
Place Eugene Bataillon, F-34095 Montpellier, Cedex 5, France}

\emailAdd{ s.davidson@lupm.in2p3.fr}

\abstract{
Lepton Flavour Violation(LFV)  is  New Physics that must  occur,  but  is
stringently constrained by experiments searching for $\mu\leftrightarrow e$
flavour change, such as $\meg$, $\meee$ or $\mec$.  
However, in an Effective Field Theory(EFT)  parametrisation,  there
are many more $\mu\leftrightarrow e$ operators than  restrictive constraints,
so  determining operator coefficients from data is a remote dream. 
It is nonetheless interesting to learn about New Physics from data,
so this manuscript introduces
 ``observable-vectors''
in the space of operator coefficients,
which  identify at any scale the
combination of coefficients probed by the  observable. These
vectors have an overlap $\gsim 10^{-3}$ with most of the coefficients,
and are
used to study whether
$\meg$, $\meee$ and $\mec$
give complementary information
about New Physics. 
The appendix
gives updated sensitivities of these processes, (and
a subset of $\tau \to \ell$  decays), to
operator coefficients  at  the weak scale  in the  SMEFT and
in the EFT  below $m_W$.
}

\begin{document} 
\maketitle
\flushbottom

%%%%%%%%%%%%%%%%%%%%%%%%%%%%%%%%%%%%%%%%%%%%%%%%%%%%%%%%%%%%%%%%%%%%%%
%% INTRO        %%%%%%%%%%%%%%%%%%%%%%%%%%%%%%%%%%%%%%%%%%%%%%%%%%%%%%
%%%%%%%%%%%%%%%%%%%%%%%%%%%%%%%%%%%%%%%%%%%%%%%%%%%%%%%%%%%%%%%%%%%%%%
%\newpage

\section{Introduction}

The observed pattern of neutrino masses implies
new particles and interactions (New Physics $\equiv$ NP)
 in the  lepton sector (beyond the Standard Model
 with left-handed neutrinos),  so one of the
 best-motivated challenges in particle physics is to
 discover what it is. Knowing where to search
 would be a useful input to this enterprise. 
One possibility would to look everywhere where NP is not
already excluded; 
another would be to explore regions suggested
by motivated models.
However, the  guidance from models can be  ambiguous,
because they are legion, and for a given process,
many models may predict
rates just beyond the  current experimental reach.
So  this manuscript  takes the agnostic view that it is
interesting to look for Lepton Flavour Violation (LFV)
everywhere it is not   excluded.
%The current aim is to   identify where that is;  once NPis detected, reconstructing  it  becomes interesting. 

This manuscript  focuses on   NP that changes lepton flavour
$\mu \leftrightarrow e$ (for a review, see {\it eg}\cite{KO}),
for simplicity
restricted to   processes  that
are otherwise flavour diagonal, such as 
$\meee$ or $\megg$,  but not  $K\to \mu^\pm e^\mp$.
In order to look for LFV ``everywhere that it is not
excluded''  among such processes,
it  would be interesting to   list observables  such that:
\ben
\item
if  $\mu \leftrightarrow e$  flavour-changing NP exists,
it  would contribute to  at least one of the %selected set of
processes. 
That is, the   set of observables   is ``complete''.

\item the observables are complementary, in the sense that
they give independent  information about the NP,
and cannot be predicted one from the other. In particular,
we want to avoid  searching for a branching ratio
that is already excluded by the upper bound on another
observable.
\een
Instead of trying to construct such a
list,  
we ask a more pragmatic question: 
to what degree are
 $\meg$,
$\meee$ and $\mec$ on nuclei
($\muc$)
complete and complementary?
These processes are selected because
the current  experimental  bounds on  the
branching ratios are restrictive
($\lsim 10^{-12}$) \cite{TheMEG:2016wtm,Bellgardt:1987du,Bertl:2006up}, and
experiments under construction aim to
improve the sensitivities to $10^{-14} \to 10^{-16}$
\cite{Mu3e,COMET,mu2e,MEGII} or better \cite{PP}.

Ideally, a   complete list  of complementary processes should be 
independent  of the theoretical formalism
used   to establish it  --- for instance,
it should not  apply   only to some models,
or depend on a choice of operator basis in  an Effective Field Theory
(EFT). 
Nonetheless,  let us start by restricting to
``heavy''  NP models,   where the
new particle masses are at a scale
$\Lambda_{NP} \gsim m_W$.
 Such models  can be parametrised
below  $\Lambda_{NP}$  in an EFT framework,
which allows to separate ``known physics'' (the data
and the Standard Model), from the theoretical
speculations above $\LNP$.
So  LFV is described 
by  operators  constructed out of  Standard Model(SM) fields
and respecting SM gauge
symmetries\footnote{This  overlooks the
implications of possible of
light NP in other sectors;  for instance a light axion/ALP or DM particle. }.
In addition, $\LNP$ is  assumed large enough to
justify retaining only a few terms in
the $ 1/\LNP^n$ expansion, which in the case of LFV,
starts at dimension six  $\propto 1/\LNP^2$. 
In this EFT context, a complete list of observables
should contain at least as many members as  there
are operator coefficients --- otherwise there
can be combinations of coefficients that are
not probed  (sometimes called ``flat directions'').
Unfortunately,  for $\mu\to e$ flavour change, 
there are many flat directions:   current
data on   $\meg$, $\meee$ and $\mec$ impose
12 to 14 \cite{DKY} bounds on the coefficients
of the 80$\to$90 operators listed in  section \ref{ssec:ops}.

A first step,
is nonetheless  to explore whether  $\meg$,
$\meee$ and $\mec$ are sensitive to
the coefficient of each operator  in
the basis of section \ref{ssec:ops}.
Such a study is  basis-dependent, and  corresponds to
calculating ``one-operator-at-a-time'' bounds,
or ``sensitivities''(the word used in this manuscript):
a coefficient smaller than such values
could not have been seen. 
This is consistent with the original aim of
identifying  where LFV is not excluded,
and the  results are  given  in the tables of
   Appendix \ref{app:sens}.
However, the coefficients can be larger than
these sensitivities, by sitting along 
various flat directions, which
are discussed in section  \ref{ssec:flatdir}.

Loop effects are the backbone of this discussion.
This is because
the contact interactions  which are induced at  tree
level in models, may   not mediate the
processes which are stringently constrained
by experiment.
Consider, for instance,  a NP model that induces a tree-level 
contact interaction $(\overline{b} \g^\a b)(\overline{e} \g_\a \mu)$.
This mediates the decay
$\Upsilon\to \overline{e} \mu$\cite{CLEOUpsilon1S}, and  contributes at one-loop
to $\meee$. But $\meee$ will have  better sensitivity,
both because  the  bound on the BR is more restrictive,
and  because  the muon lifetime is longer,
since it decays via the weak
interactions, whereas the $\Upsilon$ decays
electromagnetically. 
SM loop corrections to the contact interactions,
are discussed in more detail in section
\ref{ssec:RGEs}.

Constraints on LFV contact interactions have been tabulated
in many previous publications. This manuscript
differs from earlier works, such as \cite{LQ,bil,Carpentier,Dib,AC},
which mostly listed   tree-level bounds,
that is, constraints at the experimental
scale on the coefficients of operators which contribute at
tree level to the process. The
sensitivities tabulated here are very similar to
those of \cite{PSI},  which apply to
coefficients at a short-distance scale $\sim 1/m_W$
and  include  similar Standard Model loop corrections between
$m_W$ and low energy;  the results here complete
those of   \cite{PSI} (where
some operators are missing from the tables), extend them
by a subset of $\tau \leftrightarrow \ell$ operators, and 
also give the sensitivities  in the SMEFT basis.

The ``observable-vectors'' are defined  in Section  \ref{sec:complement}.
When New Physics is parametrised via operators in EFT,
these are constructed along theoretical
guidelines,
and identify the default basis of coefficient space.
Observable-vectors could be  an alternative basis,
corresponding to    the directions in coefficient
space  probed by observables. They
appear  implicitly in the results of
section \ref{sec:complete}. 
Their misalignment
quantifies the complementarity of
the observables:  if the vectors are orthogonal, the
observables are very complementary, if the vectors
are parrallel, the observables probe the same thing, 
so  section  \ref{sec:complement}
uses this perspective to plot the
complementarity of $\meg$, $\meee$ and $\mec$.

%%%%%%%%%%%%%%%%%%%%%%%%%%%%%%%%%%%%%%%%%%%%%%%%%%%%%%%%%%%%%%%%%%%%%%
%% SECT  %%%%%%%%%%%%%%%%%%%%%%%%%%%%%%%%%%%%%%%%%%%%%%%%%%%%%
%%%%%%%%%%%%%%%%%%%%%%%%%%%%%%%%%%%%%%%%%%%%%%%%%%%%%%%%%%%%%%%%%%%%%%

\section{Completeness}
\label{sec:complete}

Ideally,  a list of  processes  that is ``complete''
could include processes at any accessible energy scale, and
would allow to probe any   LFV contact interaction.
However   in practise,  it is difficult to search for
processes with  more than four  legs, and many restrictive
constraints come from low-energy processes. So  we  attempt
to construct a list of processes that is sensitive to all
 LFV contact interactions at  ``low'' energies $\ll m_W$
 that have three or four legs.
Section \ref{ssec:ops} lists a QCD$\times$QED-invariant
basis of operators constructed with
three or four lepton, quark, photon or gluon fields, that
change lepton flavour $\mu \to e$, and are otherwise flavour-diagonal.
A quark   and gluon basis  is used because they
appear in loops between  $m_W$ and  $ \sim 2$ GeV.
However, they live inside hadrons;
the additional step of matching quark operators onto
nucleon operators is discussed   with
$\mec$ in section \ref{ssec:brs}.
Section \ref{ssec:brs} gives the Branching Ratios for restrictively-bounded
processes in terms of the operator coefficients at the
experimental  scale,  section \ref{ssec:RGEs} describes 
how SM loop corrections are included  in this analysis,
and the branching ratios are given in terms of
weak-scale coefficients in section \ref{ssec:BRs}.
 Allowing one coefficient at a time to be non-zero in these formulae,
 gives the ``one-operator-at-a-time'' bounds, or  sensitivities,
 of  each process to each operator,
 which are collected in the  tables in   Appendix \ref{app:sens}.
 The operators can  be matched onto the SMEFT at $m_W$,
 sensitivities are also given in this  basis.
Finally, since there are more operators than experimental
constraints, section \ref{ssec:flatdir} discusses
combinations of coefficients that
are not constrained.

%This means that  our discussion will not apply to NP models that only induce  LFV contact interactions with  more than four legs (below $m_W$). 

\subsection{Operators}
\label{ssec:ops}

There are ninety  operators  which are  QCD$\times$QED
invariant, have  three or four legs, and  
which   change flavour $\mu \to e$ (and  involve no other
flavour change).  These are
suitable for describing $\mu \to e$ interactions
at  energies below  $m_W$, where the Higgs
and SU(2) bosons are not present as  external legs.
The operators  here are constructed with
chiral fermions, because  this facilitates matching onto
 the chiral SMEFT operators at the weak scale, and because
 in the lepton sector, the electron from muon decays
 is relativistic, so $\approx$ chiral, implying that
negligeable interference between operators involving $e_L$  vs $e_R$.
 The operators are added to the Lagrangian as
\beq
{\cal L} = {\cal L}_{SM} + \sum_\zeta \sum_{Lor}\frac{C^\zeta_{Lor}}{v^2}
{\cal O}_{Lor}^\zeta
+ h.c.
\label{L}
\eeq
where $1/v^2 = 2\sqrt{2} G_F$ ($v\simeq m_t$),
the operator  subscript  $Lor$ gives the Lorentz structure and
chirality of the fermion bilinears, and  the superscript $\zeta$
gives the flavour
indices. Since this manuscript only considers $\mu \to e$
transitions, all the operators contain $\bar{e}$ and $\mu$
(the $\mu^+ \to e^+$  processes are described by the $+ h.c.$),
and the $e\mu$ indices are suppressed from the superscript.

The 22 four-lepton operators are:
%Lepton flavour-changing interactions below $m_W$  can be parametrised
%as contact interactions involving quarks, leptons  photons and gluons.
%{\it aside: why   are quarks  better than hadrons? Something about
%operator dimension/number of legs? $0\nu2\b$ $ is high-dim op in nucleon theory?} 
\bea
 {\cal O}^{  ll}_{V,YY} = (\overline{e} \gamma^\a P_Y \mu ) 
(\overline{l} \gamma_\a P_Y l )  ,&&
{\cal O}^{  ll}_{V,YX} = (\overline{e} \gamma^\a P_Y \mu ) 
(\overline{l} \gamma_\a P_X l )  \nonumber \\
{\cal O}^{  ll}_{S,YY} = (\overline{e}  P_Y \mu ) 
(\overline{l} P_Y l )~~~~~~~ &&
{\cal O}^{  \tau \tau}_{S,YX} = (\overline{e}  P_Y \mu ) 
(\overline{\tau} P_X \tau )
 \label{opstauWa} \\
{\cal O}^{  \tau \tau}_{T,YY} = (\overline{e} \sigma^{\a \b} P_Y \mu ) 
(\overline{\tau} \sigma_{\a \b} P_Y \tau )~~~&& 
\nonumber
\eea
where $l\in \{e,\mu,\tau\}$,  $X,Y \in \{L,R\}$, and $X\neq Y$.
Then there are 50   operators withtwo leptons and a quark bilinear:
\bea
{\cal O}^{  qq}_{V,YY} = (\overline{e} \gamma^\a P_Y \mu ) 
(\overline{q} \gamma_\a P_Y q )~~~~~ ,&&
{\cal O}^{  qq}_{V,YX} = (\overline{e} \gamma^\a P_Y \mu ) 
(\overline{q} \gamma_\a P_X q )  \nonumber \\
{\cal O}^{  qq}_{S,YY} = (\overline{e}  P_Y \mu ) 
(\overline{q} P_Y q )~~~~~~~~~~~ ,&  &
{\cal O}^{  qq}_{S,YX} = (\overline{e}  P_Y \mu ) 
(\overline{q} P_X q )  \nonumber \\
{\cal O}^{  qq}_{T,YY} = (\overline{e} \sigma^{\a \b} P_Y \mu ) 
(\overline{q} \sigma_{\a \b} P_Y q ) ~~~~~~~~ & &
\label{2l2q}
\eea
where $q\in\{u,d,s,c,b\}$. %,  $X,Y \in \{L,R\}$, and $X\neq Y$.
And finally, there are 18 operators with  two leptons, which
include the dipoles  and operators with two photons
or gluons
\bea
{\cal O}_{D,L } = m_\mu 
 \overline{e_R} \sigma^{\a\b} \mu_L F_{\a\b}
 ~~~~&&  m_\mu  \overline{e_L} \sigma^{\a\b} \mu_R F_{\a\b}\nonumber \\
{\cal O}_{GG,Y} = \frac{1}{v}(\overline{e}  P_Y \mu ) G_{\a \b} G^{\a\b}  ~~
~~,&& {\cal O}_{G\tilde{G},Y} = \frac{1}{v}
(\overline{e}  P_Y \mu ) G_{\a \b} \widetilde{G}^{\a\b} 
\nonumber\\
{\cal O}_{GGV,Y} = \frac{1}{v^2}(\overline{e} \g_\s P_Y \mu ) G_{\a \b} \partial_\b G^{\a\s} ~~
,&& {\cal O}_{G\tilde{G}V,Y} = \frac{1}{v^2}
(\overline{e} \g_\s P_Y \mu ) G_{\a \b}\partial_\b \widetilde{G}^{\a\s} 
\nonumber\\
{\cal O}_{FF,Y} =
\frac{1}{v} (\overline{e}  P_Y \mu ) F_{\a \b} F^{\a\b} 
~~~~~~~~~,&& {\cal O}^{ }_{F\tilde{F},Y} = 
\frac{1}{v}(\overline{e}  P_Y \mu ) F_{\a \b} \widetilde{F}^{\a\b} \nonumber \\
 {\cal O}_{FFV,Y} =
\frac{1}{v} (\overline{e}  \g^\s P_Y \mu ) F^{\a \b} \partial_\b F_{\a\s} 
~~,&& {\cal O}_{F\tilde{F}V,Y} = 
\frac{1}{v}(\overline{e} \g^\s P_Y \mu ) F^{\a \b} \partial_\b\widetilde{F}_{\a\s} 
\label{obsops}
\eea
where $\sigma ^{\a\b} = \frac{i}{2}[\g^\a,\g^b]$, $X,Y \in \{L,R\}$, and $X\neq Y$.
The running of the muon mass in the dipole operators is
neglected here.
The  dimension  seven operators $O_{GG,L} , O_{GG,R}$
were  included in  $\mec$ in  \cite{CKOT},
and the other gluon operators will not be further
considered here.  The $\megg$ rate due to 
 the  various two-photon operators
 was calculated  in \cite{BCLM}, and  
the Crystal Box experiment\cite{CrystalBox}  set the constraint
$BR(\megg) \leq 7.2\times 10^{-11}$. The  $ {\cal O}_{FF,Y}$
operators  also contribute to $\mec$, and
the SINDRUMII search for $\mu Au \to e Au$  currently
has the best  sensitivity to  $ C_{FF,L}$ and $ C_{FF,R}$ \cite{DKUY}.

\subsection{Experimental bounds}
\label{ssec:brs}

We now want to relate the 
coefficients of these operators to experimental decay rates.
We restrict to the bounds on  $\meg$, $\meee$
and $\muc$ because  the current  experimental upper bounds
$BR \lsim 10^{-12}$ \cite{TheMEG:2016wtm,Bertl:2006up,Bellgardt:1987du}
are restrictive, and will improve by orders of
magnitude in coming years\cite{Mu3e,COMET,mu2e,MEGII}. 
Furthermore, the branching ratios compare to
a weak decay, so $BR \lsim 10^{-12}$ probes a new
physics scale $\Lambda_{NP} \lsim 100$ TeV.

The  Branching Ratio for $\meg$ \cite{KO}, and
the current experimental bound \cite{TheMEG:2016wtm} are
\beq
BR(\meg) = 384 \pi^2 (|C_{D L}|^2 + |C_{D R}|^2) <4.2\times 10^{-13}
\label{BRmeg}
\eeq
 so
 the dipole coefficients at the experimental scale
 should be inside the
 circle in  coefficient space given by
  eqn (\ref{BRmeg}),  that is,
  should separately satisfy $C_{D,X} \leq 1.05\times 10^{-8} $.

The branching ratio  for $\meee$ is \cite{KO,Okadameee}
\bea
BR(\meee)  &\leq& 10^{-12}\nonumber\\
& =&   2| C_{V,LL}  + 4eC_{D ,R}|^2
+2 |C_{V,RR}  + 4eC_{D, L}|^2
\nonumber\\
& &
+ |C_{V,RL}  + 4eC_{D, L}|^2
 + |C_{V,LR}  + 4eC_{D, R} |^2 
\label{BRmeee}\\
&&+ \frac{|C_{S,LL}|^2+ |C_{S,RR}|^2}{8} + (64 \ln\frac{m_\mu}{m_e} -136) 
(|eC_{D ,R}|^2 +|eC_{D ,L}|^2)
%\\ &\leq&  10^{-12}
\nonumber
\eea
where   $  \ln\frac{m_\mu}{m_e} =5.35$,
so $e^2 (64 \ln\frac{m_\mu}{m_e} -136)\simeq 204.8e^2 \simeq 18.78$.
Measuring  the polarisation of the muon  and
the angular distribution of the electrons\cite{Okadameee},
(and  even the polarisation of the electrons),
could allow to discriminate among these  various
contributions.

Combining  the $\meee$ and $\meg$ bounds in a covariance
matrix allows to obtain  separate constraints on
the dipole and vector  coefficients
(see \cite{DKY}). Since 
 the current  bound on $BR(\meg)$ is restrictive,
this amounts to imposing the bound (\ref{BRmeg})
on the dipole coefficients, and then
neglecting them in  (\ref{BRmeee}):
\bea
|C^{ee}_{V,XX}|  &\leq   &7.0\times 10^{-7}~~~,~~~
|C^{ee}_{V,XY}|  \leq  10^{-6}  \nonumber\\
|C^{ee}_{S,XX}| &\leq   &2.8\times 10^{-6}
\label{Bdmeee}
\eea
where the coefficients are evaluated at the experimental
scale. 

The conversion of a muon to an electron in
nuclei is 
a sensitive probe of $\mu \to e$ flavour
change  in the presence of quarks.
The $\mu^-$ is captured into the $1s$
state of the nucleus, and can then convert to
an electron by interacting with the
nucleons or electric field of the nucleus.
The SINDRUMII experiment at PSI, with a
continuous muon beam, 
searched for $\mec$ on Titanium
and Gold \cite{Bertl:2006up}, setting bounds
$BR(\mu A\to e +A) \lsim 10^{-12}$.
The theoretical rates for  (Spin Independent) conversion  on
many  targets  are  given in
\cite{KKO}, and  can
be written\cite{DKY}:
\beq
BR_{SI} (\mu A \to eA) =
\frac{\Gamma (\mu A \to eA)}{ \Gamma_{cap}(A)}=
B_A {\Big[}
|\hat{v}_A\cdot \vec{C}_{L} |^2 +  |\hat{v}_A\cdot \vec{C}_{R} |^2  {\Big ]}
\leq \left\{\begin{array}{ll}
 4.3\times 10^{-12} & {\rm Ti}\\
  7\times 10^{-13} & {\rm Au}
  \end{array} \right.
~~~,
\label{ip2}
\eeq
where the Branching Ratio is normalised to
the  capture rate $\mu A \to  \nu_\mu A'$
on the same nucleus, and is expressed
in terms of the coefficients $\{\tilde{C}\}$  of  operators
constructed with  a nucleons. Several comments:
\ben
\item The coefficient subscript
gives the Lorentz structure ($V$ or $S$), then the chiral projector
of the lepton current, but the nucleon current
is  not chiral because its more useful in the non-relativistic
limit to use a scalar($S$), pseudoscalar($P$),
vector ($V$), axial vector ($A$) and tensor ($T$) basis,
where the $P,A,T$  components  contribute to the Spin
Dependent rate \cite{CDK,DKS} so are neglected here.
The non-chiral coefficients can be written in terms
of chiral coefficients as, {\it eg}
\beq
C^{ ff}_{V,Y} = \frac{1}{2}(C^{ ff}_{V,YR} + C^{ ff}_{V,YL})
~~~,~~~
C^{ ff}_{A,Y} = \frac{1}{2}(C^{ ff}_{V,YR} - C^{ ff}_{V,YL})
\eeq

\item The formula  after the last
equality  is given   in the normalisation of \cite{DKY},
where
the  coefficients of nucleon operators
have been assembled in  vectors
\beq
\vec{C}_{L} = (\widetilde{C}_{D,R}, \widetilde{C}_{S,R}^{pp}, \widetilde{C}_{V,L}^{pp},\widetilde{C}_{S,R}^{nn}, \widetilde{C}_{V,L}^{nn})
\label{vecC}
\eeq
(and similarly for $\vec{C}_R$),
 and the overlap integrals  of Kitano, Koike and Okada \cite{KKO}
for target $A$ have been  assembled in
unit-normalised ``target vectors''
\bea
\vec{v}_{A}& = &(\frac{D_A}{4}, {S}_{A}^{(p)}, {V}_{A}^{(p)}, {S}_{A}^{(n)}, V_A^{(n)})
\nonumber
\eea
whose normalisation is absorbed into  the
$B_A =  \frac{32G_F^2 m_\mu^5 |\vec{v}_A|^2}{\Gamma_{cap}(A)}$.

The  vectors and normalisation factors
for Titanium and Gold are
\bea
 \hat{v}_{Ti}& =& %\left\{ \begin{array}{lr}
  (0.250,0.426, 0.458 ,0.503,0.541)  ~~~,~~~ B_{Ti} = 250
  \nonumber\\
    \hat{v}_{Au}& =& %\left\{ \begin{array}{lr}
   (0.222,0.289, 0.458 ,0.432 ,0.686) ~~~,~~~ B_{Au} = 300~~~.
 \nonumber%\\   \hat{v}_{Al}& =& (0.266 ,0.455 ,0.473 ,0.490, 0.508)~~~,~~~B_{Al} = 142 ~~~.
\label{vecv}
\eea
Since these vectors are misaligned, 
Titanium and  Gold  can measure 
independent combinations of coefficients \cite{DKY}.

\een

\subsubsection{Translating $\mec$ bounds onto quark operators}

\begin{table}[th]
\begin{center}
\begin{tabular}{|l|l|l|}
\hline
$G^{p,u}_V = G^{n,d}_V = 2$ &$ G^{p,d}_V = G^{n,u}_V = 1$&$ 
G^{p,s}_V = G^{n,s}_V = 0$ 
\\
\hline
$G^{p,u}_S = \frac{m_p}{m_u} 0.021(2) =9.0$&
$G^{p,d}_S = \frac{m_p}{m_d}  0.041(3) = 8.2$ %&   \\ 
&$  G^{p,s}_S =    \frac{m_N}{m_s}  0.043(11) = 0.42$ 
\\
 $G^{n,u}_S = \frac{m_n}{m_u} 0.019(2)= 8.1$
&$G^{n,d}_S = \frac{m_n}{m_d} 0.045(3)= 9.0$&
$G^{n,s}_S =    \frac{m_N}{m_s}  0.043(11) = 0.42$
\\
$G^{N,c}_S = \frac{2m_N}{27m_c} $
&$G^{N,b}_S = \frac{2m_N}{27m_b}$&  \\
\hline
\end{tabular}
\caption{\label{tab:G} This table is taken from \cite{DKS}.
It gives matching coefficients between  nucleon
  and light-quark-flavour-diagonal  operators. The  parenthese
  gives the uncertainty in the last figure(s).
 The  %second
$G^{N,q}_S$  were obtained via EFT methods\cite{Hoferichter:2015dsa,Alarcon:2011zs}, and  an average of lattice results~\cite{Junnarkar:2013ac}
is used for the strange quark. The heavy quark scalar $G_S$s
 are from \cite{SVZ}.
In all cases, the 
 $\overline{\rm MS}$  quark masses  at $\mu = 2$~GeV
 are taken as  $m_u = 2.2$~MeV,   $m_d = 4.7$~MeV, and
 $m_s = 96$~MeV~\cite{Agashe:2014kda}.
 The nucleon masses are  $m_p =  938 ~{\rm MeV}$
 and $m_n =  939.6 ~{\rm MeV}$.
}
\end{center}
\end{table}

The nucleon coefficients can be written in terms
of the quark coefficients (that we wish to constrain)
using the expectation values of  quark-currents
in the nucleus $\{G_\Gamma^{N,q}\}$, defined as, {\it eg}
$\langle N| \bar{q}(x) q(x)| N\rangle = G_{S}^{N,q} \bar{u}_N(P_f)  u_N (P_i) e^{-i (P_f- P_i) x}$,  and given in  table \ref{tab:G}.
For  the vector and scalar coefficients:
\bea
\widetilde{C}_{V,Y}^{NN} & =& \sum_{q\in u,d,s} G^{N, q}_V C_{S,Y}^{qq} \nonumber\\
\widetilde{C}_{S,Y}^{NN} & =& \sum_{q\in u,d,s} G^{N, q}_S C_{S,Y}^{qq}
+ \sum_Q G^{N, Q}_S C_{S,Y}^{QQ}  -\frac{8\pi m_N}{9\a_s m_t } C_{GG,Y}
\label{matchscal}
\eea
where 
 $Q\in \{c,b\}$. For the scalars, the first term is  the
tree contribution  of light quarks in the nucleon,
the second is  the one-loop contribution of heavy quarks to
the gluon density (which  contributes to the
scalar nucleon current), so the explicit gluon contribution
given in the last term only contains  contributions
from $m_W$ or above (such as the  $(\overline{t} t)(\bar{e}\mu)$ operator).

This allows to translate the upper bound on the
Branching ratio on Gold $BR_{Au}$ to  a bound  on the quark-level coefficients
at a scale  $\sim$  2 GeV: 
\bea
4.9 \times 10^{-8} &= &
\sqrt{ \frac{BR^{exp}_{Au}}{B_{Au}}} \nonumber\\
&\gsim & {\Big|}
0.222 C_{D,R}  +
1.60 C^{uu}_{V,L} +  1.83 C^{dd}_{V,L}
+ 6.10 C^{uu}_{S,R} +
6.258 C^{dd}_{S,R}
\label{bdmec2} \\&&
 + 0.303 C^{ss}_{S,R} + 0.721\frac{2m_N}{27m_c}C^{cc}_{S,R}
+ 0.721\frac{2m_N}{27m_b}C^{bb}_{S,R}
 -0.721 \frac{8\pi m_N}{9\a_s v } C_{GG,R} 
{\Big|} \nonumber
\eea
where $v\simeq m_t$, and $C^{bb}_{S,R}$ is at $m_b$ rather than 2 GeV. 
The same bound applies for $L \leftrightarrow R$.

The SINDRUMII upper bound on the Branching ratio on Titanium,
(see eqn \ref{ip2}),  is slightly less sensitive to
individual operator coefficients, but  constrains
a different linear combination. In the formulation
of \cite{DKY},  a target nucleus of charge $Z$  can be viewed as unit  vector
$\hat{v}_Z$ in the space of operator coefficients,
such that $BR \propto | \hat{v}_Z \cdot  \vec{C}|^2$.
This allows to write the direction probed
by Titanium as the direction probed by Gold
plus an orthogonal vector:
$$
\hat{v}_{Ti} =  \hat{v}_{Au} \cos\phi +  \hat{v}_{\perp}  \sin\phi
$$
where $ \cos\phi = \hat{v}_{Ti} \cdot \hat{v}_{Au}$. So one can
solve for $\hat{v}_{\perp} $, and with a covariance matrix obtain
that  the  SINDRUM bound on Titanium  gives a constraint
on  $| \hat{v}_\perp \cdot  \vec{C}| $:
\bea
\sqrt{ \frac{BR^{exp}_{Ti}}{B_{Ti} \sin^2\phi}}&= &
6.24\times 10^{-7}\gsim  {\Big|} 0.155 C_{D,R}  
-0.506 C^{uu}_{V,L} -1.168 C^{dd}_{V,L} \label{bdmec2b} \\&&
+ 9.21 C^{uu}_{S,R} +
9.02 C^{dd}_{S,R}
+ 0.446 C^{ss}_{S,R}
\nonumber\\ &&
+ 1.062\frac{2m_N}{27m_c}C^{cc}_{S,R}
+ 1.062\frac{2m_N}{27m_b}C^{bb}_{S,R}
 -1.062 \frac{8\pi m_N}{9\a_s v } C_{GG,R}
{\Big|} \nonumber
\eea
and the same bound applies for $L \leftrightarrow R$.

\subsection{Accounting for loop effects}
\label{ssec:RGEs}

In a Wilsonian sense, 
the coefficients  in the branching ratios of the previous section are 
evaluated  at  a  scale $\Lambda_{low}$ near the experimental scale.  In
order to estimate  sensitivities to 
coefficients at $\LNP$, 
 loop corrections  on scales $\Lambda_{low}\to \LNP$
 must be  ``peeled off''.
These loops  corrections are evaluated here in an EFT perspective,  where
only SM particles are dynamical below $\LNP$, and new particle propagators
are expanded  in $p^2/\LNP^2$  around a contact interaction\footnote{This
 allows to implement  Wilsonian intuition  about
Renormalisation Group running of operator coefficients, while
computing loops in dimensional regularisation.}.
These SM loops  are then independent of the
New Physics, and only need to be calculated once
as Renormalisation Group running of operator coefficients,
and possibly as matching coefficients as discussed below. 
As usual with the Renormalisation Group, a one loop
calculation generates  the   $(\log/16\pi^2)^n$ terms for all $n$,
a 2 loop calculation
generates the  $(\log^{n-1}/(16\pi^2)^{n}) $ terms, and so on.
Provided that   $\LNP$  is sufficiently large,  only
a few terms in the $1/\LNP^2$  expansion are required.

The log-enhanced loops  included here arise from the one-loop
Renormalisation Group Equations (RGEs)
of QED and QCD, which  can be written in matrix form as
\beq
\mu \frac{d }{d \mu} \vec{C} =  
\frac{\alpha_{s} (\mu)}{4\pi} \vec{C} {\Gamma}^{s} +  
\frac{\alpha_{e}}{4\pi}\vec{C} {\Gamma}  ~~~,
\label{RGE}
\eeq 
where the operator coefficients are lined up in
the row vector ${\vec{C}}$.  
The diagonal anomalous dimension matrix  ${\Gamma}^{s}$ of QCD
only renormalises  the scalar and tensor operators involving
 a quark bilinear. Since
 the QCD coupling $\alpha_s$  is large and runs significantly,
 its one-loop effects are resummed :$C(\mu_f) =
(\frac{\alpha_s(\mu_f)}{\alpha_s(\mu_i)})^{\gamma/2\b_0}
C(\mu_i)$\cite{burashouches}.
Two-loop QCD effects are neglected
--- although this gives an uncertainty of ${\cal O}(10\%)$ ---
because  QCD is less interesting than QED:  QED
mixes operators, allowing to transform an operator
that is difficult to probe experimentally, into one
that is tightly constrained. 
Finally,  ${\Gamma}$ is
the well-populated  anomalous dimension matrix of QED,
augmented by  two-loop
QED mixing\footnote{The two-loop effects are included in
the 'Hooft-Veltman scheme for $\g_5$, where the one-loop matching contributions
vanish, so should give a scheme-independent result.} of vector operators
into the dipole\cite{PSI}.

The RGEs are solved analytically as an expansion in
$\a_e$, which is approximated not to run;
  we retain the ${\cal O}(\alpha_e \log)$ terms, and some of the
 ${\cal O}(\alpha_e^2 \log^2)$ and ${\cal O}(\alpha_e^2 \log)$ terms.
 The aim is to include  effects that are $\gsim 10^{-3}$. 
The solution is  formally a scale-ordered exponential,
which can be expanded
in $\alpha_e$ by defining $dt = d\mu/\mu$,
and substituting $\frac{\a_s}{4\pi}\Gamma^s = -iH_0$,
$\frac{\a_e}{4\pi}\Gamma^T = -iH_{int}$, into the
solution for  the time-translation operator given in
chapter 4 of \cite{P+S}.  At  ${\cal O}(\a_e)$, this gives:
\bea
 \vec{C}(\mu_f)&\simeq &  \vec{C}(\mu_i  )D_s\left[ I  +
 \frac{\a_e}{4\pi}   \widetilde{\Gamma} \ln \frac{\mu_f}{\mu_i} +...
\right] \nonumber \\
\widetilde{\Gamma}_{KJ}& = &\frac{ \Gamma^e_{KJ}(\mu_f)}{1 +a_J - a_K -a_d} 
\frac { 1 - \lambda^{a_J-a_K -a_d+1 } }{1-\lambda}
\equiv f_{KJ}\Gamma^e_{KJ}
\label{soln1}
\eea
where there is no sum on $KJ$
in $f_{KJ}\Gamma^e_{KJ}$, $D_s =$diag$\{1,...\lambda^{-a_{S,T}}\}$
describes the diagonal QCD running of scalar or tensor operators
involving quarks  with $\lambda = \frac{\a_s(\mu_f)}{\a_s(\mu_i)}$,
$a_T= -4/23$, $a_S= 12/23$,  $a_J = 0$ for $J\neq S,T$, and $a_d$
describes the running of $\Gamma^e_{KJ}$  in the case where
it includes a running QCD parameter (eg $a_d = a_S$ for anomalous dimensions mixing
quark tensor operators to the dipole).
At ${\cal O}(\a_e^2)$,
the scale-ordered exponential
gives $\lsim 10\%$ QCD corrections to the 
mixing of scalars to the dipole (``Barr-Zee''), so
 these QCD corrections are  neglected
in the analytic solutions given later.

The particle content of the EFT changes  when the scale
crosses  a particle mass, so the operator
basis changes too. At each such threshhold, the
Greens functions in the theory above and below
must be identical, which allows to match
coefficients from the basis above onto below. 
(Matching at the weak scale  is recently discussed in \cite{DS}.)
Frequently, this matching can be
performed at tree level in both theories; 
in some cases, the leading
contribution of
an operator arises via
loop diagrams of the theory above,
in which case these are included
(hopefully after checking that they
are scheme-independent, which is
not always the case\cite{Roma}).
So for instance,  the one-loop matching
of scalar $b,c$ quark operators onto the gluon operators,
at the quark mass scale, 
 is included as given in eqn (\ref{matchscal}).

What level of accuracy  is required?  The aim is to  include
the dominant contribution of each  operator (at $\LNP$)
to each process. Since almost every operator
contributes to each observable
(see the tables of  appendix  \ref{app:sens}),
the question is whether the largest
contributions have been included. This is difficult
to demonstrate, because the dominant contribution may not be
the lowest  order in every perturbative expansion
(loops, couplings, scale ratios); an  example is the
two-loop Barr-Zee
diagrams, which  give the dominant
contribution of flavour-changing Higgs couplings
to radiative decays like $\meg$.
However, the selection of terms included here has
been checked against various models\cite{2hdm,meg@mW}.
Notice also that  including  effects $\gsim 10^{-3}$
does not mean the calculation is accurate to  three
figures; rather, the resulting constraints
on operator coefficients have uncertainties
$\gsim 10\%$. 

Missing  from the results given here, are most   strong
interaction effects beyond  Leading Order, below 2 GeV. 
For example,  quark loop contributions to $\meg$ and $\meee$
are included at scales above 2 GeV, but should
be replaced by pion loops or resonances from
2 GeV to $m_\m$. An interesting study in
this direction is \cite{aneesh}.
The $\mec$ calculation is also
at Leading Order in $\chi$PT.

\subsection{Constraints at $m_W$}
\label{ssec:BRs}

This section gives the experimental
constraints of eqns (\ref{BRmeg}),(\ref{Bdmeee})
and (\ref{ip2}) expressed in terms of
operator coefficients at the weak scale.

\subsubsection{$\meg$}

The MEG  bound on   coefficients  at $m_W$ is
\bea
 1.05\times 10^{-8} &\gsim& 
{\Big | } C_{D,X}\left(1- 16 \frac{\alpha_e}{4\pi} \ln\frac{m_W}{m_\mu}\right)
\nonumber \\
&&- \frac{\alpha_e}{4\pi e}  \left( -8
   \frac{m_\tau}{m_\mu}C^{\tau \tau}_{T,XX} \ln\frac{m_W}{m_\tau} 
 +  C^{\mu \mu}_{S,XX} \ln\frac{m_W}{m_\mu}
 +  C^{ee}_{S,XX} \frac{m_e}{m_\mu}\ln\frac{m_W}{m_\mu} 
+C_{2loop} \ln\frac{m_W}{m_\tau}
\right)
\nonumber \\
&&+ 8 \frac{\alpha^2_e}{e(4\pi)^2} \ln^2\frac{m_W}{m_\tau} \left(
  \frac{m_\tau}{m_\mu}C^{\tau \tau}_{S,XX} 
 \right)
\nonumber \\
&& -8 \lambda^{a_T} \frac{\alpha_e}{4\pi e}\ln\frac{m_W}{2~{\rm GeV}} \left( 
 -\frac{m_s}{m_\mu}C^{ss}_{T,XX}
 +2 \frac{m_c}{m_\mu}C^{cc}_{T,XX}
 -\frac{m_b}{m_\mu}C^{bb}_{T,XX} 
  \right)f_{TD} 
\nonumber \\
&&+ 8 \frac{\alpha^2_e}{3e(4\pi)^2} \ln^2\frac{m_W}{2~{\rm GeV}} \left(
\sum_{u,c} 4 \frac{m_q}{m_\mu}C^{qq}_{S,XX}
+ \sum_{d,s,b} \frac{m_q}{m_\mu}C^{qq}_{S,XX}
 \right)
 {\Big | }~~~~ 
 \label{megbdmW}
\eea
where all the coefficients  are to be evaluated
at $m_W$, but the masses are on-shell
($=m_\psi(m_\psi)$).
Several comments are in order:
\ben
\item At one loop,   all the  tensor operators,
  and the scalars  involving
  3 muons or electrons,
  can mix to the dipole
 by closing
same-flavour legs and attaching a photon. 
  (However, the $3e$ scalar coefficient can be
neglected, because $\meee$ sets a more stringent bound on this operator.)

\ben
\item 
The $\tau$-tensor is in the second parenthese, the quark
tensors are in the fourth (the light quark tensors are neglected,
because they contribute at one loop to SI $\mec$, unsuppressed
by $m_q/m_\mu$).
QCD  causes  $C^{qq}_{T,XX}$ to run,  and also  the quark
mass   that appears in the anomalous
dimension mixing  the tensor to the dipole is
taken running. These QCD effects  are
described by $\lambda = \alpha_s(2{\rm GeV})/\alpha_s(m_W) \simeq 2.18$,
and 
\bea
f_{TD} =\left(   \frac{1}{1-a_T-a_S}  \frac{1 - \lambda^{1-a_T -a_S}}{1 - \lambda}  \right)
\eea
where   $a_T = -4/23$, $a_S = 12/23$ are
respectively the anomalous dimensions of tensors and scalars/masses in QCD,
and $\lambda^{a_T} \simeq 0.873$ and $ f_{TD} \simeq 0.862$. In eqn (\ref{megbdmW}), the quark masses are at low energy: $m_q(m_q)$.

\item It is interesting that the coefficients of the
dipole and tensor operators can be comparable,
allowing for unexpected cancellations. Numerically,
the bounds of eqn(\ref{megbdmW})
is
\bea
\!\!\! C_{D,X} (m_\mu) &\gsim& 
{\Big | } 0.938 C_{D,X}(m_W)
+ 0.981 C^{\tau \tau}_{T,XX} (m_W)   -0.75 C^{cc}_{T,XX} +...
 {\Big | }~~~~ ~~~~
 \label{TDcancel}
\eea
\een

\item 
$C_{2loop}$ is a combination of vector operator coefficients
that mix to the dipole via the 2-loop RGEs\cite{Roma}
\bea
C_{2loop} &=& -\frac{\alpha_e}{4\pi}\left[
 \frac{58}{9}C^{\tau\tau}_{V,YY}
 +  \frac{116}{9}\sum_{l=e,\m}C^{ll}_{V,YY}
+ \frac{64}{9}(C^{uu}_{V,YY} + C^{cc}_{V,YY})  +
 \frac{22}{9}\sum_{q=d,s,b}C^{qq}_{V,YY}
\right. \nonumber\\
&&\left. 
-\frac{80}{9}(C^{uu}_{V,YX} + C^{cc}_{V,YX})
- \frac{14}{9}\sum_{q=d,s,b}C^{qq}_{V,YX}
- \frac{50}{9}\sum_{l=e,\m,\tau}C^{ll}_{V,YX}
\right. \nonumber\\ &&\left.
+ 4\sum _{f=b,c,s,\tau} C_{S, YX}^{ff} \frac{Q_f^2 N_f m_f}{m_\mu}
\right]
\label{C2loop}
\eea
where the chirality of the  electron outgoing from these operators must
be the same as that of the dipole operator into which they mix
---so the combination above mix into $C_{D,X}$ for $X\neq Y$.

The  $C_{2loop}$ term  in eqn (\ref{megbdmW})  is compact but
not quite correct: in reality, the lower cuttoff of the logarithm
should be $m_\tau$ for the $\tau\tau$ operators, $m_b$ for the $bb$ operators,
2 GeV for the $cc$,$ss$,$uu$ and $dd$ operators, and 
$m_\mu$ for the $\mu\mu$  and $ee$ operators. 

\item the  $\tau$ and (heavy) quark scalar operators mix to the tensor
at one loop, and therefore the dipole at two-loop. Since the tensor-to-dipole
mixing is ${\cal O}(1)$ for heavy fermions because  enhanced
by the mass, the scalar-dipole mixing is included for the $\tau$ on
the third parenthese, and  for the heavy quarks in the last
parenthese. The QCD running
  is $\lsim 10\%$, if the heavy  quark
masses are taken at low energy $m_Q(m_Q)$, so neglected.
The lower cutoff of the logarithm for quark operators
is approximated as 2 GeV  to given a shorter formula,
but should be  chosen as
a function of the quark flavour ---eg  $m_b$ for the $b$ quark,
and 2 GeV for $u,d,s$.

\een

\subsubsection{$\meee$}

The decay $\meee$  constrains  the magnitude of several
combinations of coefficients, as given in eqn (\ref{BRmeee}).
At one-loop, there are no operators that mix into
the scalar operators  ${\cal O}^{ ee}_{S,...}$.
All  vector operators ${\cal O}^{ ff}_{V,XY}$ and ${\cal O}^{ ff}_{V,XX}$
can mix to  ${\cal O}^{ ee}_{V,XR}+{\cal O}^{ ee}_{V,XL}$
by closing the $f$ legs and attaching a photon to the loop, then
attaching the photon to $e^+e^-$. These ``penguin'' diagrams
mix the combination
\beq
Q_e C_{ping1,X} = Q_e \frac{4}{3}\sum_f N_f Q_f (C_{V,XL}^{ff} +  C_{V,XR}^{ff})
~~~~~f\in\{e,\m,\tau,u,d,s,c,b  \}
\label{ping}
\eeq
into both  $C^{ee}_{V,XR}$ and $ C^{ee}_{V,XL}$.   The $3e$ and $3\mu$
vector operators also mix to  the $3e$ vectors via a second
 penguin diagram which closes a different selection of legs,
 and the $3e$ operators are renormalised by photon exchange
 between the legs. 
 As a result,   the  constraints  from  $\meee$ (see eqn \ref{Bdmeee}),
 combined with the MEG bounds on $\meg$,
 give the following bounds on coefficients
 evaluated at $m_W$:
 \bea
 7\times 10^{-7}  &>&
 C^{ee}_{V,XX} \left( 1-12 \frac{\a_e}{4\pi} \ln\frac{m_W}{m_\mu} \right)
-\frac{\a_e}{4\pi} \left(
 \frac{4}{3} \left(  C^{\mu\mu}_{V,XX}  +  C^{ee}_{V,XX} \right)
 \ln\frac{m_W}{m_\mu} +  C_{ping1,X}\ln\frac{m_W}{m_\tau}  \right)
 \nonumber \\
 10^{-6}&>&
  C^{ee}_{V,XY}\left( 1+12 \frac{\a_e}{4\pi} \ln\frac{m_W}{m_\mu} \right)
-\frac{\a_e}{4\pi}  \left(
 \frac{4}{3} \left(C^{\mu\mu}_{V,XX} +  C^{ee}_{V,XX} \right)\ln\frac{m_W}{m_\mu}  
+  C_{ping1,X}  \ln\frac{m_W}{m_\tau}  \right)
 \nonumber \\
2.8 \times 10^{-6} &>&  C^{ee}_{S,XX} \left( 1+12 \frac{\a_e}{4\pi} \ln\frac{m_W}{m_\mu} \right)
\label{meeepeng1} 
\eea
where $X\in \{L,R\}$ and $X\neq Y$. The  overall logarithm
multiplying $ C_{ping1,X} $  is  an approximation to
shorten the formula;  more correctly,  the 
 $bb$ coefficients  in $ C_{ping1,X} $ should be
multiplied  by a logarithm cut off at %$\ln\frac{m_W}{m_b}$
 $m_b$,
and the logarithm for the $ee$ or $\m\m$ coefficients
should end at $m_\mu$ (the correct logs are
implemented in the numerical bounds in the Appendix). (And as usual, there
should be hadronic loops to replace the light quarks
below  2 GeV.)

\subsubsection{$\mec$}

The  SINDRUMII bound on $\mu$Au$\to e+$Au, given in eqn (\ref{ip2})
for out-going left-handed electrons, 
can be expressed as  the bound   on coefficients at $m_W$:
\bea
4.9 \times 10^{-8}&\gsim&  {\Big|}\left(
0.222 C_{D,R}(m_\mu) + 1.602C^{uu}_{V,L} + 1.830C^{dd}_{V,L} -0.721 f_N^{GG}\frac{8\pi m_N}{9\a_sm_t}C_{GG,R}\right)
\nonumber\\
&&{ +\lambda^{a_S} {\Big (}
1+  \frac{ \alpha}{ 4\pi}  \frac{26}{3} \widetilde{\ln}  {\Big )} }
 \left[  6.10C^{uu}_{S,R} +\frac{1.30m_N}{27m_c}C^{cc}_{S,R}
  \right]  \nonumber\\
&&   \qquad  \qquad 
{ +\lambda^{a_S} {\Big (}
1+  \frac{ \alpha}{ 4\pi}  \frac{20}{3} \widetilde{\ln}  {\Big )} }
\left[ 6.26C^{dd}_{S,R} +0.303 C^{ss}_{S,R} 
 +  \frac{1.30m_N}{27m_b}C^{bb}_{S,R}\right]  \nonumber \\
% &&+ 2 \lambda^{a_S}   \frac{ \alpha}{ 4\pi} \widetilde{\ln} {\Big [} 6.1 C^{uu}_{S,R}   + \frac{1.30m_N}{27m_c} C^{cc}_{S,R} {\Big ]}  \nonumber \\
&&
- 16 f_{TS}  \lambda^{a_T} \frac{ \alpha}{ 4\pi} \widetilde{\ln}
\left( 2[6.100 C^{uu}_{T,RR}   + \frac{1.30m_N}{27m_c} C^{cc}_{T,RR} ]
\right. \nonumber\\
&&\left. \qquad  \qquad
- 6.258 C^{dd}_{T,RR} - 0.303C^{ss}_{T,RR}  -\frac{1.30m_N}{27m_b}C^{bb}_{T,RR} \right) 
\label{bdmecAumW} \\
&&
-\frac{\alpha}{2\pi} \widetilde{\ln}
{\Big (} 6.41C^{uu}_{A,L}  { -}  3.66C^{dd}_{A,L}
- 0.305[{ C^{ee}_{V,L} } +  C^{\m\m}_{V,L} {- C^{ee }_{A,L}}- C^{\mu \mu}_{A,L}] 
+ 0.228 C_{ping,1}  {\Big )}  {\Big|}
\nonumber
\eea
where $\widetilde{\ln}= \ln(m_W/m_{low}$)  should be inside
the square brackets because $m_{low} \in\{m_b, 2~{\rm GeV},m_\mu\}$
depends on the coefficient, the masses are
on shell ($m_\psi(m_\psi)$) and the dipole contribution
is given at the experimental
scale (it can be written in terms of coefficients at $m_W$
with eqn  \ref{megbdmW}). A similar bound applies with
$L$ and $R$ interchanged. 

The bound from Titanium on an orthogonal combination of
coefficients, given in eqn  (\ref{bdmec2b}), can be
written with the same conventions as
\bea
\sqrt{ \frac{BR^{exp}_{Ti}}{B_{Ti} \sin^2\phi}}&= &
6.24\times 10^{-7}\gsim  {\Big|} \left(0.155 C_{D,R}  
-0.506 C^{uu}_{V,L} -1.17 C^{dd}_{V,L}
-1.06 \frac{8\pi m_N}{9\a_s v } C_{GG,R} \right)
\nonumber \\&& %
{ +\lambda^{a_S} {\Big (}
1+  \frac{ \alpha}{ 4\pi}  \frac{26}{3} \widetilde{\ln}  {\Big )} }
\left[9.21 C^{uu}_{S,R}  + \frac{2.12m_N}{27m_c}C^{cc}_{S,R}\right]
 \label{bdmecTimW} \\ &&
 \qquad \qquad {
+\lambda^{a_S} {\Big (} 1+    \frac{ \alpha}{ 4\pi} \frac{20}{3} \widetilde{\ln}  {\Big ) } }
\left[9.02 C^{dd}_{S,R} + 0.446 C^{ss}_{S,R} + \frac{2.12m_N}{27m_b}C^{bb}_{S,R}
\right]   \nonumber \\%OK
&&
- 16 f_{TS}  \lambda^{a_T} \frac{ \alpha}{ 4\pi}  \widetilde{\ln}
\left(  2[9.21 C^{uu}_{T,RR}   +\frac{2.12m_N}{27m_c} C^{cc}_{T,RR} ]
\right.\nonumber\\
&&\left. 
\qquad \qquad- 9.02 C^{dd}_{T,RR} - 0.446C^{ss}_{T,RR}  -\frac{2.12m_N}{27m_b}C^{bb}_{T,RR} \right) 
\nonumber\\
&&
-\frac{\alpha}{2\pi} \widetilde{\ln}
{\Big (} 2.576C^{dd}_{A,L}  -2.024C^{uu}_{A,L} 
- 0.0346[ C^{\m\m}_{V,L} - C^{\mu \mu}_{A,L}] 
+ 0.32 C_{ping,1}  {\Big )}%
{\Big|} \nonumber
\eea
where $\widetilde{\ln}$ is defined
after eqn (\ref{bdmecAumW})   and the masses are
on shell ($m_\psi(m_\psi)$).

\subsection{What is not probed?}
\label{ssec:flatdir}

This section gives an incomplete  list  of  coefficient combinations
that  are  not   probed  by $\meg$, $\meee$, and
$\mec$.  There are 45 $\mu \to e_L$ operator coefficients
 in section \ref{ssec:ops}, upon which
the  current bounds on $\meg$, $\meee$, $\mu {\rm Au} \to e_L{\rm Au}$
 and $\mu {\rm Ti} \to e_L{\rm Ti}$ set 1+3+1+1 = 6 constraints \cite{DKY}.
Including $\megg$, which was searched for by
Crystal Box \cite{CrystalBox} would give an  additional  constraint (on a
$\g\g$ operator)\cite{BCLM,DKUY}, but there remain $\lsim$
40  unconstrained directions in coefficient space.
The case of operators
that produce an outgoing $e_R$ is the same and independent,  because
 the Branching Ratios independently constrain
 both processes.

This section does not give a list of three (or six) dozen flat directions. 
The  subset of  flat directions listed here are
selected because they  are  relatively
stable under RG running, or  ``natural'',
according to the notion 
 introduced it \cite{DKY}
(for cancellations among operator coefficients).
It assumes that   model parameters
at  the New Physics scale $\Lambda_{NP}$ do not
know the experimental scale,  so coefficients should
not cancel against logarithms of scale ratios --- because the
scale ratio  could be  chosen by the observer  but
the coefficient is determined by the theory.

An elegant way to implement this notion of
naturalness, is to only admit cancellations
which are RG-invariant. 
For instance, if one  restricts to  operators containing
a quark bilinear, and only considers their
one-loop diagonal QCD running, then cancellations
among  the   vector coefficients, or among the scalars, 
or the tensors, are RG-invariant.
This could be generalised to include 
the one-loop  QCD and QED running, by 
diagonalising the full  anomalous dimension matrix.
However,  the diagonalisation
would be difficult,  and the eigenvectors  would be
curious combinations of  different Lorentz structures
and  external legs --- as opposed to the
usual ``intuitive''  basis of section \ref{ssec:ops},
where the operators correspond to
potentially distinguishable interactions of  physical
particles. Furthermore, the
only degenerate eigenvalues might be
among vector operators involving fermions of
the same electric charge (because scalars
mix to tensors whose mixing to
the dipole is proportional to the fermion mass). 

A more pragmatic implementation of this
notion of ``natural'' cancellations is to use the
perturbative solution of the RGEs given
in eqn (\ref{soln1}),  and allow cancellations
among coefficients which  1) run with the same QCD
anomalous dimension,  and 2) multiply  a similar
$\alpha_{em}\ln \mu_f/\mu_i$ factor.
The formulae for the  constraints
on  coefficients at the weak scale,
given in eqns (\ref{megbdmW}, \ref{meeepeng1},
\ref{bdmecAumW}, and \ref{bdmecTimW}),
are arranged to  display these  ``natural''
cancellations, or flat directions:
each parenthese of the formulae satisfies
the two conditions. 
So any combination of coefficients that  causes
a parenthese to vanish is a  ``natural''
flat direction.

An example of  flatish directions, or  operators which are
poorly constrained by $\meg,\meee$ and $\mec$ are
the axial vector operators of the form
$$
(\overline{e}\g_\a P_X \mu) (\overline{f}\g^\a \g_5 f)~~~,~~~f\in\{s,c,b,\tau\}
$$
which  contribute to $\meg$  at 2-loop with
a mass enhancement
(see eqn \ref{C2loop}). There should be three
combinations of these operators, orthogonal to
eqn (\ref{C2loop}), which are ``flat'' to the order calculated here. 
(The axial vector  operators with $f \in\{ u,d,\mu\}$ are
not in the list,  because they contribute
at one QED loop to $\mec$.) % so are  less well constrained than the vectors operators.
A second example  would be
any combination of vector operators
$$
\sum_f C_f (\overline{e}\g_\a P_X \mu) (\overline{f}\g^\a  f)~~~,~~~f\in\{s,c,b,\tau\}
$$
where the $\{C_f\}$  are chosen  orthogonal to
$C_{ping,1}$ (see eqn \ref{ping}), and  also
to $C_{2loop}$ if one wishes large  $C_f$. 

Reference \cite{DKY}'s  notion of naturalness    precludes cancellations
among the combinations of coefficients in
parentheses in eqns (\ref{megbdmW}, \ref{meeepeng1},
\ref{bdmecAumW}, and \ref{bdmecTimW}),  so it could
be interpreted as
transforming the single constraint from $\meg$ into
a  constraint on each of the five parentheses, and the 2 bounds from $\mec$ into
eight.  However,   ``unnatural''  cancellations
can occur, see for instance  eqn (\ref{TDcancel}).

%%%%%%%%%%%%%%%%%%%%%%%%%%%%%%%%%%%%%%%%%%%%%%%%%%%%%%%%%%%%%%%%%%%%%%
%% SECT  %%%%%%%%%%%%%%%%%%%%%%%%%%%%%%%%%%%%%%%%%%%%%%%%%%%%%%
%%%%%%%%%%%%%%%%%%%%%%%%%%%%%%%%%%%%%%%%%%%%%%%%%%%%%%%%%%%%%%%%%%%%%%

\section{Complementarity}
\label{sec:complement}

The aim of this section is  to show that  the
observables considered here ($\meg$, $\meee$ and $\mec$)
give independent information about  models.
To do this, an alternative basis is proposed for coefficient space.
This basis is constructed from observables,
 spans the subspace they probe, and can be defined at
all scales.

In an EFT  framework, 
a model $ {\cal M}$ can be represented by the  vector $\vec{C}_{\cal M} (\Lambda)$
of operator coefficients
it induces\footnote{Since models usually have parameters, there would be  a vector
for each choice of parameters, and
varying the model parameters would scan
over the vectors  of coefficients that
can represent the model.}, which  is scale-dependent but  relatively simple to calculate  at the New Physics
scale $\Lambda_{NP}$.
An observable  can  be represented as
one or several combinations 
of  operators
whose coefficients it probes. This also depends on the scale, and   is 
 relatively simple to establish  at the experimental
scale $\Lambda_{expt}$ --- for instance,
$\mu\to e_L \gamma$ probes the %``vector'' of
operator  ${\cal O}_{D,R} (m_\mu)$. 
However, to obtain the rate for an 
observable, % from a given effective Lagrangian, %   in  model ${\cal M}$,
one must evaluate the matrix element of the operators
and  integrate phase space.  So 
it is convenient to define  observable-vectors   $\vec{v}_{obs} (\Lambda)$
to live in the vector space of the coefficients,
such that 
the rate for an observable
can be written 
\beq
\Gamma \propto \sum |\vec{C}_{\cal M} (\Lambda) \cdot \vec{v}_{ obs} (\Lambda)|^2
  \label{Cdotv}
  \eeq
at a common scale $\Lambda$. For instance, at the
experimental scale,  eqn (\ref{BRmeg}) implies that the  two
observable-vectors for $\mu\to e \gamma$ point in the  dipole directions,
so one can choose
\bea
\vec{v}_{\mu\to e_L \gamma} (m_\mu) =  \hat{u}_{D,R}~~~~ \nonumber\\
\vec{v}_{\mu\to e_R \gamma} (m_\mu) =  \hat{u}_{D,L}~~~.
\label{obsvmeg}
\eea
 where $ \hat{u}_{D,R}, \hat{u}_{D,L}$
are  unit vectors  such that $\vec{C} \cdot  \hat{u}_{D,R}=
 {C}_{D,R}$. Some  observable-vectors for $\mec$
 at the experimental scale are already given in eqn (\ref{vecv}).

By assumption, the dynamics below
$\LNP$ is Standard Model, so the model and observable
vectors  can be translated in scale
by the SM RGEs. Since models are legion and observables
are few, it might be more efficient to translate the
observable vectors to $\LNP$, rather than calculating
 $\vec{C}_{\cal M} (\Lambda_{expt})$, as is commonly done.
Indeed, if the observable-vectors were   known  as a function of
$\LNP$, then the predictions of  a New Physics model
 would be  simple to obtain from eqn (\ref{Cdotv}).
Part of the translation  $\vec{v}_{obs} (\Lambda_{expt})
\to \vec{v}_{obs} (\Lambda_{NP})$  appears   in section
\ref{sec:complete}, where  the combination
$\vec{C}_{\cal M} (m_W) \cdot \vec{v}_{\mu\to e_L \gamma} (m_W)$
  is given  by eqn (\ref{megbdmW}) with $X=R$,
$\vec{C}_{\cal M} (m_W) \cdot \vec{v}_{\mu{\rm Au} \to e_L{\rm Au}} (m_W)$
appears   in eqn
(\ref{bdmecAumW}), and the
appropriate inner product for
the orthogonal constraint from
$\mu{\rm Ti} \to e_L{\rm Ti}$
is given in eqn (\ref{bdmecTimW}).
Similar formulae apply for  outgoing $e_R$, but
the focus below is mostly on  $e_L$.
Matching onto the SMEFT, and running up to $\LNP$ is
left for a later work.

The observable vectors corresponding to $\meee$ can  also be constructed.
For this, it is convenient to take an idealised theoretical perspective,
imagining that  the chirality of the four leptons can be
observed.  Combined with the electron angular distributions
\cite{Okadameee}, this
allows to define  ``observable'' vectors for $\meee$
at the experimental scale 
\bea
\vec{v}_{\mu_L\to e_L\overline{e_L} e_L} (m_\mu) &= &
\sqrt{2}[\hat{u}^{ee}_{V,LL}  + 4e \hat{u}_{D,R}]
\nonumber\\
\vec{v}_{\mu_R\to e_L\overline{e_R} e_L} (m_\mu)& =& \frac{1}{2\sqrt{2}}\hat{u}^{ee}_{S,LL} 
\nonumber\\
\vec{v}_{\mu_R\to e_R\overline{e_L} e_L} (m_\mu) &= &
\hat{u}^{ee}_{V,RL} (m_\mu) + 4e \hat{u}_{D,R} 
\eea
plus another three  vectors with $ L \leftrightarrow R$, and the dipoles. 
These vectors can be expressed at $m_W$ using eqns
(\ref{megbdmW}) and (\ref{meeepeng1}).

 These vectors corresponding to $\meg$, $\meee$ and $\muc$,
 allow to quantify the complementarity of these processes at
 any scale. If, at  the  chosen scale,  
  the vectors   remain  ``relatively orthogonal'',  then the observables
 are complementary. If the vectors  are aligned, then
 the observables become identical probes of New Physics.
It is interesting to study   complementarity  at $\LNP$,
because
 we are looking for  information about the New Physics. 
However,  as an illustrative  exercise, 
we start  at the experimental scale.

Notice that the complementarity of observables
is a theoretical question, unrelated to the
magnitude of current experimental bounds.
In this  it differs from the  correlation matrix
 that can be constructed from the experimental constraints,
which  defines the  allowed ellipse in coefficient space;
the direction and magnitude of the axes of the ellipse
depend on the relative magnitude of the constraints. 
Complementarity  is also
 model-independent, and unrelated to
  correlations between observables that could arise
 in some models. Such correlations would occur
 if the projections
of $\vec{C}_{\cal M}$ onto the observable vectors have similar
scaling with the parameters of model ${\cal M}$ :
$$
\frac{\vec{C}_{\cal M} \cdot \vec{v}_{obs1}}{\vec{C}_{\cal M} \cdot \vec{v}_{obs2}}
={\rm independent~of~model~parmeters} ~~~.
$$
The notion of ``model discriminating power'', present in the literature
\cite{CKOT,Celis:2014asa, Giffels:2008ar}, is related to complementarity,
but  compares the projection of different observable vectors, onto
different model vectors.
The discriminating power of observables therefore depends on the
models considered, whereas  their complementarity  is model-independent.

At the  experimental scale, $\meg$, $\meee$ and $\mec$
have different external particles, so  naively appear complementary.
However,
the dipole contributes to all three processes,
preventing the observable-vectors from being
orthogonal even at the experimental scale.
Indeed,   at $\Lambda\sim m_\mu$, $C_{D,R}(m_\mu)$ and $C_{V,LL}(m_\mu)$ contribute
about equally  to $\mu_L\to  e_L\overline{e_L} e_L$,
so $\vec{v}_{\mu_L\to e_L\overline{e_L} e_L} (m_\mu)$
is misaligned   with respect to $\vec{v}_{\mu_R\to e_L \gamma} (m_\mu)$
by $\sim \pi/4$:
\bea
\frac{\vec{v}_{\mu_L\to e_L\overline{e_L} e_L}  \cdot \vec{v}_{\mu_R\to e_L \gamma}}
{|\vec{v}_{\mu_R\to e_L \gamma} ||\vec{v}_{\mu_L\to e_L\overline{e_L} e_L} |} (m_\mu)
%\simeq .77
\sim \cos (\pi/4) ~~~.
\eea
The dipole also contributes to
$\mec$, but with a smaller weight (see eqn \ref{bdmec2}),  such
$\vec{v}_{\mu_R\to e_L \gamma}$(2 GeV)  and
  $\vec{v}_{\mu{\rm Au} \to e_L{\rm Au}}$(2 GeV)
 are almost orthogonal (separated by
 an angle of $\sim 88$ degrees).
 So  at low energy, an  approximately orthogonal basis 
for the  observable-vector subspace  can be constructed 
with the two dipoles, and the remaining observable-vectors with
the dipole subtracted. For instance,
\bea
\vec{u}_{\mu \to e_L \gamma}(\Lambda) &=&
 \vec{v}_{\mu \to e_L \gamma}(\Lambda)  \nonumber\\
 \vec{u}_{\mu{\rm Au} \to e_L{\rm Au}}(2~{\rm GeV})
 &=&
  \vec{v}_{\mu{\rm Au} \to e_L{\rm Au}}(2 ~{\rm GeV})
-0.222  \vec{v}_{\mu_R\to e_L \gamma}(2 ~{\rm GeV}) \nonumber\\
\vec{u}_{\mu_L\to e_L\overline{e_L} e_L} (m_\mu) &= &
\vec{v}_{\mu_L\to e_L\overline{e_L} e_L} (m_\mu) -
4\sqrt{2}e \vec{v}_{\mu_R\to e_L \gamma}(m_\mu)
\label{basis}
\eea
which are called $\vec{u}$ to distinguish them from the 
observable-vectors  $\vec{v}$.

\begin{figure}[ht]
\unitlength.5mm
\begin{center}
\epsfig{file=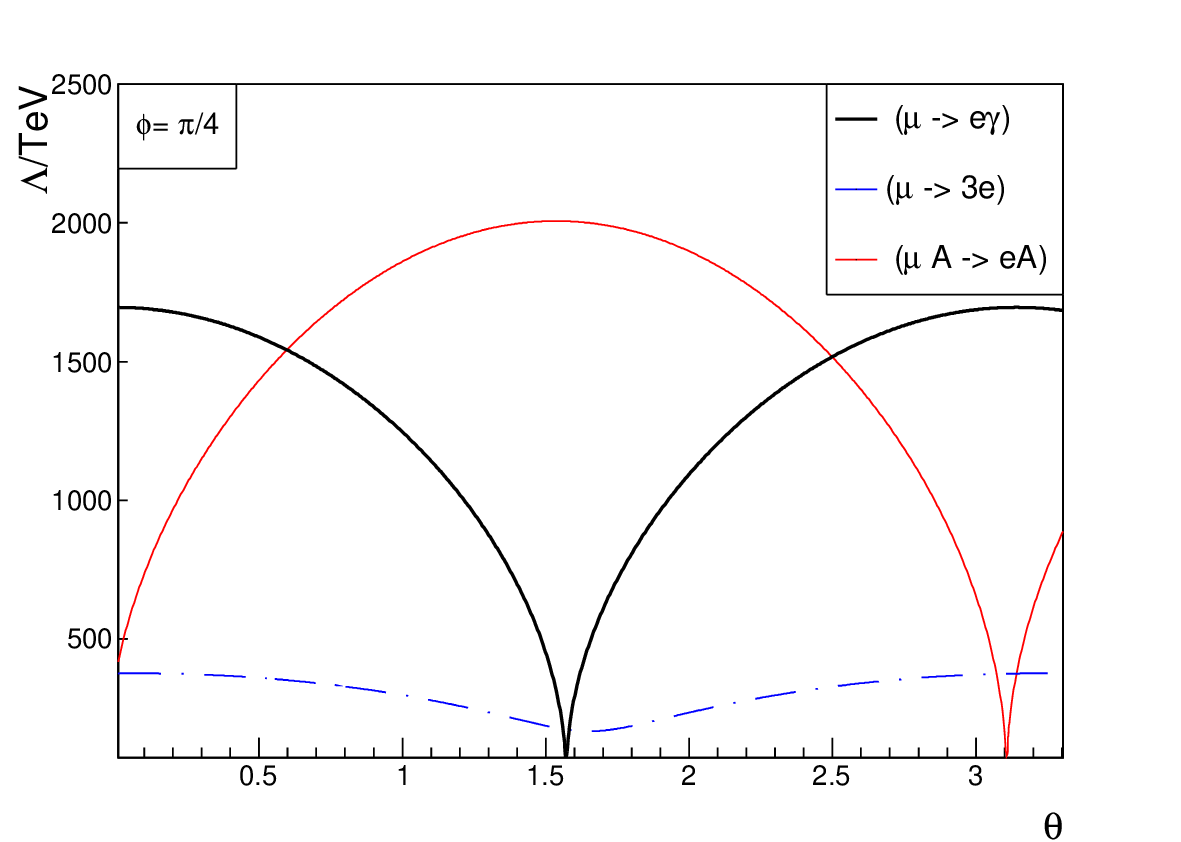,height=6cm,width=8cm}
\epsfig{file=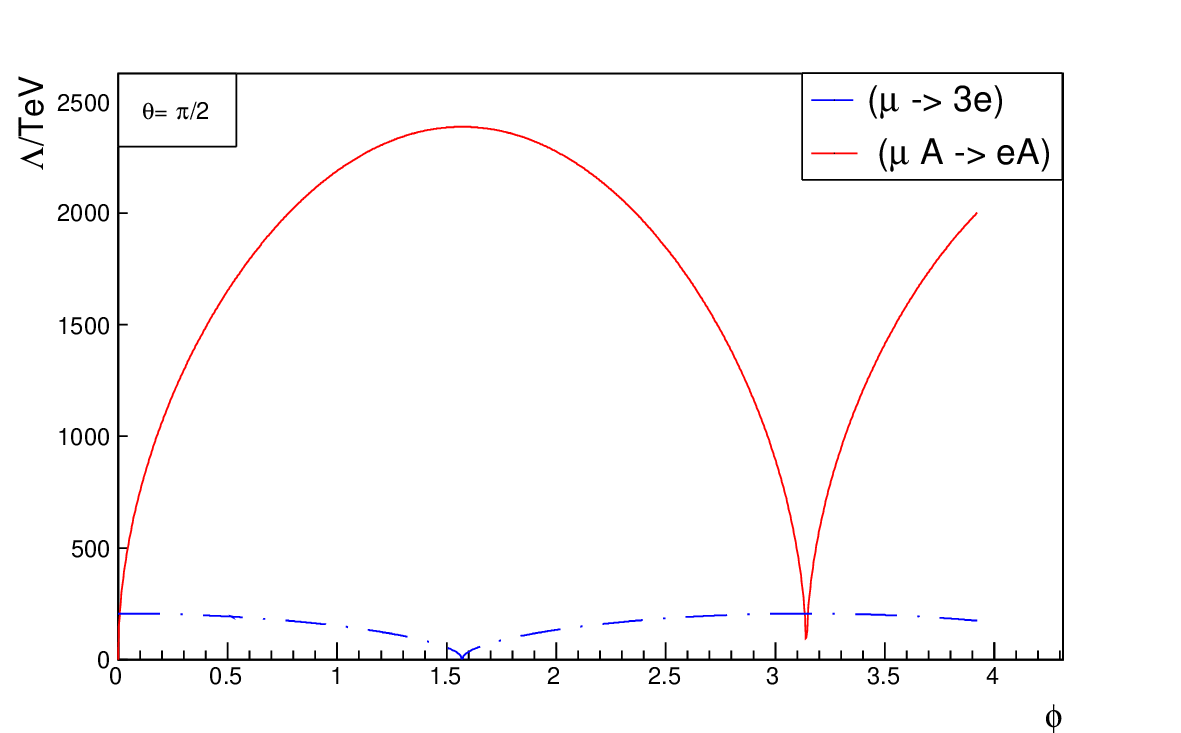,height=6cm,width=8cm}
\end{center}
\caption{ The  bound on
  the New Physics scale $\LNP$ from  current
  constraints on $\megL$ (thick black),
$\meeeL$ (dot-dashed blue) and $\mucL$ (thin red),
for New Physics parametrised  at the experimental scale
by the effective Lagrangian given in eqn (\ref{quel}).
The first figure is  as a function of $\theta$ for $\sin \phi =
\cos \phi = 1/\sqrt{2}$, and the second is as a function
of $\phi$ for $\cos \theta = 0$. 
\label{fig:expt}
%MADE BY ORDI/ROOT/COMPLMRTY/K1plotBB3.C
}
\end{figure}

This  situation at  low energy   is illustrated
in figure \ref{fig:expt}, where low energy is taken to be
2 GeV in order to use quark operators for $\mec$.
The { plots are} in polar coordinates in the subspace probed by 
$\mu \to e_L \gamma$, ${\mu{\rm Au} \to e_L{\rm Au}}$
and $\mu \to e_L\overline{e_L} e_L$, where
the vertical $z$-axis corresponds to
the dipole $\vec{u}_{\mu \to e_L \gamma}$(2 GeV),
and the $x$  and $y$ axes respectively
to  $\vec{u}_{\mu_L\to  e_L\overline{e_L} e_L}$(2 GeV) and
$\vec{u}_{\mu {\rm Au} \to e_L{\rm Au}}$(2 GeV).
{ This subspace corresponds to a Lagrangian at the experimental scale of 
\bea
{\cal L}_{eff}% (2~{\rm GeV})
(m_\mu) =\frac{1}{\LNP^2}
{\Big [} c_ \theta m_\mu\overline{e}\sigma \cdot F P_R \mu+
  s_\theta c_ \phi (\overline{e}\gamma^\a P_L\mu)(\overline{e}\g_\a P_Le)
  +s_\theta s_ \phi{\cal O}_{Au} {\Big ]} + h.c.
\label{quel}
\eea             }
   where 
  ${\cal O}_{Au} $ is the  combination of
   quark  operators probed  on Gold, 
   $s_\theta = \sin \theta$, $c_\phi= \cos \phi$ and so on,
   { and the radial coordinate is taken $|\vec{C}| = v^2/\LNP^2$,
     so that
     $$
 2\sqrt{2} G_F |\vec{C}|^2 =\frac{1}{\LNP^2}~~~.
     $$}
This is the complete low-energy Lagrangian for
the considered observables; the only ``toy'' aspect is
that   chirality of the electrons  is fixed  to make the
parameter space of plottable dimension.

{ The  resulting  BRs are:
\bea
BR(\megL) &=& 384 \pi^2 |\vec{v}_{\megL}\cdot \vec{C}|^2 =  384 \pi^2 |\vec{C}|^2 |\cos \theta|^2 \nonumber\\
BR(\meeeL) &=&   |\vec{v}_{\mu_L\to  e_L\overline{e_L} e_L}\cdot \vec{C}|^2
+ 18.76 |\vec{v}_{\megL}\cdot \vec{C}|^2   \nonumber\\
&=&  |\vec{C}|^2( 2|\sin \theta \cos \phi  + 1.2  \cos \theta |^2
+ 18.76  |\cos \theta|^2) \label{toplot} \\
BR(\mu {\rm Au} \to e {\rm Au}) &=& 300 | \vec{v}_{\mucL} \cdot \vec{C} |^2
= 300 |\vec{C}|^2 |0.222 \cos \theta + 9.08 \sin \theta \sin \phi  |^2 ~~.   \nonumber
\eea
Setting the BRs equal to their experimental  limits (given
in section \ref{ssec:brs}),
gives a bound on  $|\vec{C}|$ which is plotted
in  figure \ref{fig:expt} as a function of $\theta,\phi$. 
The angular dependence of the bounds  illustrates
that the observables  are
the complementary at the experimental scale --- which
was already obvious.}
The BR$(\meeeL)$ vanishes at
$\phi=\pi/2$ in the { second} plot (where
$\theta = \pi/2$) because  both  the dipole
and four-fermion contributions vanish;  the
BR does not quite  vanish at $\theta= \pi/2$
in the { first}  plot because the subdominant four-fermion
contribution is present at $\phi = \pi/4$.
One also sees that  $\mec$ gives the best bound on $\LNP$,
for comparable coefficients of all three operators. This
is because the conversion process is coherent on
the nucleus (so is enhanced at large  atomic number),
and because the scalar quark densities  in the nucleon
 are large ($\sim 9$, see table \ref{tab:G}---
the Branching Ratio due to a vector quark
current would be $\sim 10\to 100$ smaller).

Consider now the complementarity of  these observables
at $m_W$ ($m_W$ is used as  a simple, although inadequate, substitute for
$\LNP$). In their evolution from the
experimental scale to the weak scale,
the  observable-vectors
will rotate in coefficient space and change
length (the
non-unitary  matrix  that evolves coefficients in scale
can be written as
a rotation, a diagonal  matrix, and another rotation).
The degree of orthogonality between the vectors
can be obtained by taking inner products;  
at the weak scale, the basis  vectors
for the processes plotted here 
(= dipoles, plus observable-vectors with
their dipole components subtracted)
are still
 orthogonal to  three figures  despite their significant
 rotations (this is understandable; the vector space  has  $\sim 90$
 dimensions).
 But their  lengths change;
 in particular the dipole  vectors  grow by more than a factor two,
 due to the large loop contributions shown in eqn
 (\ref{megbdmW}).

\begin{figure}[ht]
\unitlength.5mm
\begin{center}
\epsfig{file=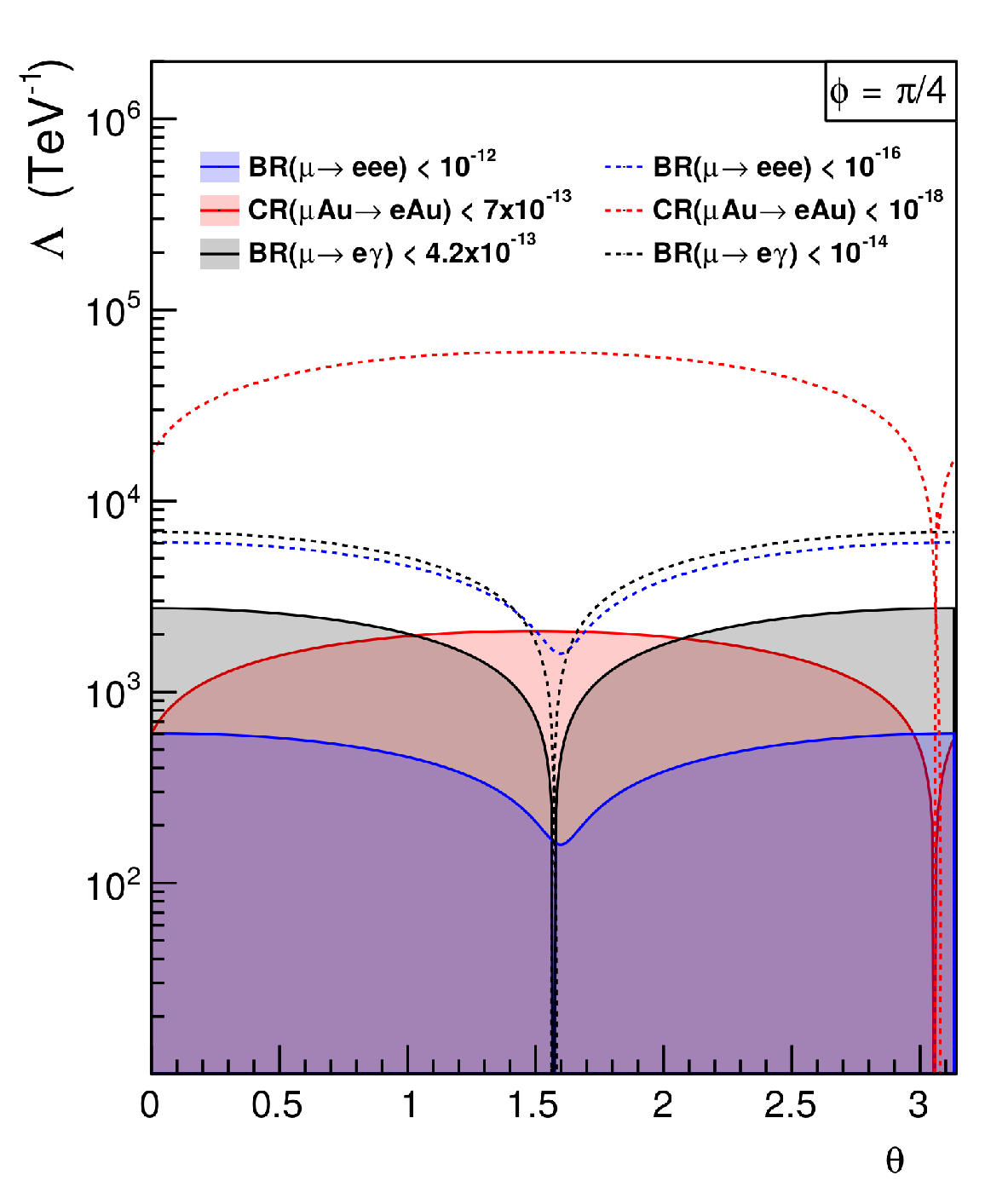,height=10cm,width=7cm}
\epsfig{file=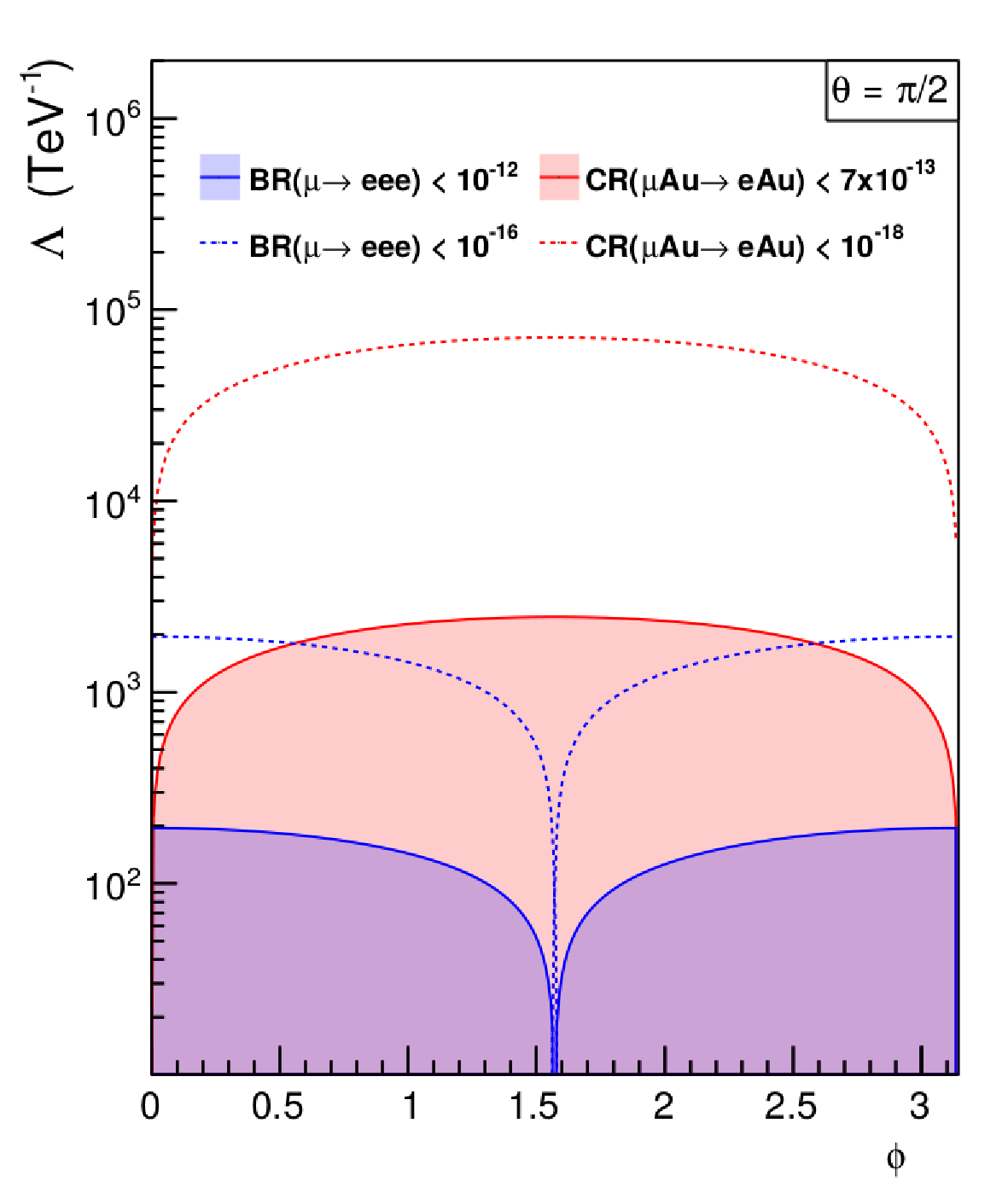,height=10cm,width=7cm}
\end{center}
\caption{ { The  bound on $\LNP$ from current
  constraints on  $\megL$ (black; vanishes at $\theta= \pi/2$),
$\meeeL$ (blue; dips at$ \theta,\phi \sim \pi/2$) and $\mucL$ (red; vanishes at$\theta,\phi \sim \pi$), as a function of
  coefficients at   $m_W$. Dotted lines
  indicate possible bounds from future experimental
  sensitivities. The  operator coefficients are parametrised in
  spherical coordinates in the subspace  probed by the
  experimental processes: the
vertical axis is the direction   probed
by $\megL$ (see eqn \ref{megbdmW}) and the  $xy$ plane is
spanned by   $\vec{u}_{\mu{\rm Au} \to e_L{\rm Au}}(m_W)$
and
$\vec{u}_{\mu_L\to e_L\overline{e_L} e_L} (m_W)$,
see eqn (\ref{basis}). The first plot is for $\phi = \pi/4$,
and the second for $\theta=\pi/2$ (where the dipole vanishes).}
See discussion at the end of section \ref{sec:complement}.
\label{fig:mW}
%MADE BY ORDI/ROOT/KAPPAPLOTS/SDTheta.C,deBEPlot.C
}
\end{figure}

 The observables
 nonetheless remain complementary at $m_W$, as shown in 
figure \ref{fig:mW}.
These  plots  are similiar to
those of  figure \ref{fig:expt},
{
  %giving the bound on $\LNP$ as a function of the angles $\theta$ and $\phi$,  obtained from  the  current experimental  constraints  on $\megL$, $\meeeL$ and $\mucL$,  (the dotted lines are  the sensitivities that could be reached by future experiments).
but
 differ in  that the operator
 coefficients are at $m_W$ (standing in for $\LNP$). 
 The first plot shows  $\LNP$ as a function of the} angle
$\theta$ between  the model vector $\vec{C}(m_W)$,
and the direction probed by the dipole $\vec{u}_{\mu\to e_L \gamma}(m_W)$
(at $\phi = \pi/4$, that is a model with identical
coefficients in the directions $\hat{u}_{\mu_L \to e_L\overline{e_L} e_L}  $
and  $\hat{u}_{\mu{\rm Au} \to e_L{\rm Au}}$). The second plot
is the projection of the model-vector onto the plane
perpendicular to the dipole.  { Were this plot
at $\LNP$, rather than $m_W$,   and if  the three 
Branching Ratios were observed,  it would allow
to extract  the all the information these
rates give  about New Physics  (via dimension six operators
and leading order  RGEs), at the New Physics scale, so without the blurring
effect of SM loops. }
As expected, $\meg$ is maximal at $\theta= 0,\pi$,
and vanishes for $\theta = \pi/2$, and
$\mec$ is larger when the model vector is
more orthogonal to the dipole. { The
$BR(\meeeL$)   follows $BR (\megL)$
due to the enhanced (by RG running) contribution of the dipole
operator to $\meeeL$;
it does not vanish with
$\vec{C}\cdot\vec{u}_{\mu\to e_L \g}$ at
$\theta= \pi/2$ due to the
four-fermion contribution.
Similiarly,
$BR(\mu{\rm Au} \to e_L{\rm Au})$
vanishes at $\theta \lsim \pi$ where
there is a cancellation between the
negative $\vec{C}\cdot\vec{u}_{\mu\to e_L \g}$
and positive
$\vec{C}\cdot\vec{u}_{\mu{\rm Au} \to e_L{\rm Au}}$.}

%%%%%%%%%%%%%%%%%%%%%%%%%%%%%%%%%%%%%%%%%%%%%%%%%%%%%%%%%%%%%%%%%%%%%%
%% SECT  %%%%%%%%%%%%%%%%%%%%%%%%%%%%%%%%%%%%%%%%%%%%%%%%%%%%%%
%%%%%%%%%%%%%%%%%%%%%%%%%%%%%%%%%%%%%%%%%%%%%%%%%%%%%%%%%%%%%%%%%%%%%%

\section{Discussion and Summary}
\label{sec:disc}

 Reconstructing  New Physics (NP)  from data is a dream for
many phenomenologists.  If New Physics  is heavy, then 
Effective Field Theory can
be a tool in pursuing  this  dream, because it
allows to separate what is known --- the Standard Model(SM) and low energy
data about NP --- from the NP that is pursued.  In particular, 
it allows theoretical  travels in energy scale,
from the experimental scale  towards the NP scale $\LNP$,
because  the dynamical degrees of freedom are by assumption
in  the SM.  However, in this manuscript,
the experimental constraints are translated only
as far as $m_W$; reaching $\LNP$ is left for later work.

  The   processes considered here
%So this manuscript sets out to discover what
%could be learned about NP from
are  $\meg$, $\meee$, and $\mec$, because
current experimental constraints are stringent
($BR \lsim 10^{-12}$) and are expected to improve
significantly in coming years ($\to 10^{-16}$).
These are low-energy processes, occuring at
an energy-transfer $\sim m_\mu$, whose
 experimental Branching Ratios (BRs)
 are reviewed in section
\ref{ssec:brs}. Together,
they set $\sim$  a dozen  constraints on
 $\mu \leftrightarrow e$ contact
 interactions at the experimental scale.

 We would like to know  what
these processes can  tell us about NP, with
as few  assumptions about the NP as possible. 
A first assumption is that the new particles are
heavy, with masses at a scale $\LNP\gsim m_W$ (possibly $\gg m_W$).
Furthermore, they are assumed to  generate
some  three or four-particle,
 $\mu \to e$
flavour-changing contact interactions
(a list is given in section \ref{ssec:ops}).

With these assumptions, section \ref{sec:complete} explores whether,
if  $\mu \leftrightarrow e$
 flavour-changing
NP  exists, we should  see  $\meg$, $\meee$ and/or $\meg$?
%This question is explored in section \ref{sec:complete}, with the above  assumptions.
In order to  reach the short-distance NP interaction
that mediates a low-energy $\mu\leftrightarrow e$
transition, the intermediate-scale SM loop corrections
must be peeled off,  for instance
using the Renormalisation Group
Equations (RGEs) summarised in section \ref{ssec:RGEs}.
This scale evolution  can  transform  one
$\mu \leftrightarrow e$
interaction  into another, 
so at short distances/high scales,
 the experimental constraints
apply to  lengthly 
combinations of coefficients. 
These are given in 
section \ref{ssec:BRs} at
the scale $m_W$, and  the
appendix tabulates  the sensitivities  to
individual operator coefficients
(one-at-a-time bounds), both in the  basis of
section \ref{ssec:ops} and in the SMEFT. 
These results include  one-loop and
some two-loop effects,  and should give  the leading
contribution of each operator to each
process
%, and contributionssuppressed by $\gsim 10^{-3}$
 (see discussion in section
\ref{ssec:RGEs}). 
The tables show that at this accuracy,  $\meg$,
$\meee$ and $\mec$ are sensitive to
(almost) all the  operators or section \ref{ssec:ops}
(the possible  exceptions
are a few  $\mu$-$e$-gluon-gluon
operators--- see eqn \ref{obsops}
and following discussion).
 Recall that the
operator list represents all QED$\times$QCD-invariant
LFV three or four-legged contact interactions, in
a chiral representation for fermions.

  It is unclear if this is  reassuring, because cancellations
in a matrix element  can occur among different coefficients.
Such  directions  in coefficient space
are sometimes refered to as  flat directions. 
There are many such flat directions
%Although  most $\mu\to e$ operators constructed with 3 or 4 SM fields do contribute to the considered  experimental processes,
 in the coefficient space for $\mu\to e$ flavour change,
after imposing the  dozen bounds from $\meg$,$\meee$ and
 $\mec$, because there are ${\cal O}(100)$ operators. 
Indeed, there are so many flat directions,
some 
one  could be  motivated to introduce a  limit
 on how much cancellation can be ``naturally''
 allowed. 
 The flat directions  are discussed in section \ref{ssec:flatdir}.

However, more interesting that the flat
directions, are the directions that the
observables do constrain, because these
are what we can use  to discriminate among models.
So section \ref{sec:complement}  introduces vectors in coefficient
space, which correspond to observables (these
are  patterned on the ``target vectors'' of
\cite{DKS,DKY}). At the  experimental scale, there
can be several observable-vectors for a
give process, each selects  
a combination of  coefficients who 
interefere in the rate:
$$
\Gamma \propto \sum  |\vec{C}_{\cal M}(\Lambda_{expt}) \cdot \vec{v}_{obs}(\Lambda_{expt})|^2
$$
where $\vec{C}_{\cal M}(\Lambda_{expt})$ is the vector
of coefficients corresponding to model ${\cal M}$,
evaluated at the experimental scale  $\Lambda_{expt}$. 
For instance,  for $\meg$, whose BR is given in eqn
(\ref{BRmeg}), the two observable-vectors at
the experimental scale are the unit vectors  that
select the coefficients of the dipole operators
in the vector $\vec{C}_{\cal M}(\Lambda_{expt})$.
The vectors for the remaining observables
are given in section \ref{sec:complement}.
The observable-vectors are interesting, because
they are scale-dependent, and can be translated
to $\LNP$ with the Renormalisation Group Equations.
Then there would be  no need to run  operator coefficients
down to the experimental scale for every model;
rather, rates could be computed from coefficients
at $\LNP$, by dotting them into $\vec{v}_{obs}(\LNP)$
This is left for future work in the lepton sector.

  In section \ref{sec:complement}, the observable
vectors are used to study the complementarity
of  experimental processes. The vector(s) corresponding
to a given observable indentifies the subspace of coefficients
that the observable probes, so  the misalignment angle between
the observable-vectors of different processes,
quantifies the complementarity of the processes.
When the vectors evolve in scale,
this misalignment angle can grow or shrink, 
indicating  that the  observables
become more, or less, complementary at high scales.
For the purpose of learning about New Physics, clearly 
it is  desirable for observables to be
complementary at $\LNP$. This manuscipt
 only reaches the weak
scale in RGE evolution,  so figure
%\ref{fig:expt} and
\ref{fig:mW}  illustrates the  complementarity of
$\meg$, $\meee$ and $\mec$ at the
scale $m_W$.

\subsection*{Acknowlegdements}

{I thank B Echenard for encouraging interest and
useful plots.}

\appendix

%\section*{Appendices}
%\label{}

%%%%%%%%%%%%%%%%%%%%%%%%%%%%%%%%%%%%%%%%%%%%%%%%%%%%%%%%%%%%%%%%%%%%%%
%%       %%%%%%%%%%%%%%%%%%%%%%%%%%%%%%%%%%%%%%%%%%%%%%%%%%%%%%
%%%%%%%%%%%%%%%%%%%%%%%%%%%%%%%%%%%%%%%%%%%%%%%%%%%%%%%%%%%%%%%%%%%%%%

\section{Tables of  Sensitivities }
\label{app:sens}

The tables in this Appendix give the sensitivities
of %$\meg$, $\meee$ and  $\mu Au \to eAu$
various processes (listed in the first row of the tables)
to the operator coefficients  given in the left column
and evaluated at $m_W$. The ``sensitivity'' of an (unobserved)
process to a coefficient is calculated
%from the  rates
%for $\meg$, $\meee$ and $\mu Au \to eAu$,
%given in section \ref{sec:complete}
by allowing  only that coefficient
to be non-zero at  $m_W$. Coefficients
 smaller than these values
are too small to have been observed,
but larger 
coefficients could be allowed,
if their contributions  
cancel  against  other  coefficients.

The tabulated results arise from the experimental bounds given
in eqns (\ref{BRmeg},\ref{BRmeee},\ref{ip2}) for muon decays, and  in table \ref{tab:BRtau} for
$\tau$ decays.

Tables \ref{tab:ls1}, \ref{tab:qs}, \ref{tab:ls1tau} and \ref{tab:qstau} contain the  QED$\times$QCD-invariant
operators  relevant below the weak scale, and  listed in section \ref{sec:complete} for $\mu\to e$ flavour change
The operators  are added to the Lagrangian with the normalisation given in
eqn (\ref{L}):
$$
{\cal L} = {\cal L}_{SM} + 2\sqrt{2} G_F  \sum  C^\zeta_{Lor}{\cal O}^\zeta_{Lor} + h.c.~~~~,~~2\sqrt{2} G_F = \frac{1}{v^2}~,~v\simeq m_t
$$
where the subscript is the Lorentz contraction, and
the superscript $\zeta$ represents the flavour indices.
Notice that in the normalisation used here,  all
operators annihilate muons (or $\tau$s) and create electrons; the
reverse process is assured by the $+ h.c$.  So
the  coefficients of an operator and its 
conjugate have the same magnitude. 

At the weak scale $m_W$, the low energy operator basis can be matched onto the
SMEFT, so the sensitivities can  be expressed in this basis.
Tables \ref{tab:2lsSMEFT} to   \ref{tab:2l2qsSMEFT}
(for $\mu\leftrightarrow e$)
and   \ref{tab:2l2qsSMEFTtau} (for $\tau\leftrightarrow e$)
apply to operators in
the SMEFT basis, added to the Lagrangian as:
\beq
{\cal L} = {\cal L}_{SM} + 2\sqrt{2} G_F  \sum C^\zeta_{J}{\cal O}^\zeta_{J} + h.c. 
\label{SMEFT}
\eeq
where the sum is over all dimension six operators and  all
flavour indices.  
In the usual SMEFT formulation, the $+h.c.$ is not included
for hermitian operators;
%({\it eg} four-fermion operators such that
%${\cal O}^{e\mu}_{Lor} = {\cal O}^{e\mu}_{Lor}$);
here  there is $+h.c.$ for all operators but
the hermitian  operators are defined with a
factor 1/2 (which gives the usual normalisation for
coefficients, because the operator and its
conjugate contribute to the Feynman rule.). The
SMEFT convention of summing all flavour indices
causes some four-lepton operators to appear several times:
\bea
 {\cal O}^{e \mu \ell\ell}_{L L} =
 {\cal O}^{\ell\ell e \mu}_{L L}  = {\cal O}^{ \ell \mu e \ell}_{L L} ={\cal O}^{e \ell\ell \mu}_{L L}  \nonumber\\
  {\cal O}^{e \mu \tau\tau}_{L L} =
 {\cal O}^{\tau\tau e \mu}_{L L}  ~~~,~~~ {\cal O}^{ \tau \mu e \tau}_{L L}
 ={\cal O}^{e \tau\tau\mu}_{L L}  \nonumber\\
 {\cal O}^{e \mu ll}_{RR} =
 {\cal O}^{ll e \mu}_{RR}  = {\cal O}^{ l \mu e l}_{RR}= {\cal O}^{e ll \mu}_{RR}
\eea
for $\ell\in\{e,\mu\}$ and  $l\in\{e,\mu,\tau\}$. So the coefficients of
these identical operators are also identical,  and the bounds in
 table \ref{tab:4lsSMEFT} apply to the appropriate sum of coefficients:
\bea
 \sum C^{e \mu \ell\ell}_{L L}&  =& C^{e \mu \ell\ell}_{L L}+
 C^{\ell\ell e \mu}_{L L}  + C^{ \ell \mu e \ell}_{L L} +C^{e \ell\ell \mu}_{L L} =
 4 C^{e \mu \ell\ell}_{L L} \nonumber\\%\label{eqLL}\\
\sum C^{e \mu \tau\tau}_{LL} &=&  C^{e \mu \tau\tau}_{L L} + C^{\tau\tau e \mu}_{L L} =
2  C^{e \mu \tau\tau}_{L L} \nonumber\\
\sum C^{ \tau \mu e \tau}_{LL} &=&  C^{ \tau \mu e \tau}_{L L} + C^{e \tau\tau\mu}_{L L} =
2  C^{e \tau\tau\mu}_{L L} \nonumber\\%\label{eqLLtau}\\
\sum C^{e \mu ll}_{RR}& =& C^{e \mu ll}_{RR}+
 C^{ll e \mu}_{RR}  +C^{ l \mu e l}_{RR} +C^{e ll \mu}_{RR}
 \label{eqsumRR}
\eea
Finally, the CKM matrix was neglected in matching, so {\it eg},
low energy operators ${\cal O}_{V,RL}^{uu}$ and  ${\cal O}_{V,RL}^{dd}$
both match onto the SMEFT operator  ${\cal O}_{EQ}^{dd}$.

Tables \ref{tab:ls1}, \ref{tab:qs}, and \ref{tab:2lsSMEFT}
to  \ref{tab:2l2qsSMEFT} give the sensitivity of
$\meg$, $\meee$ and $\mec$  to $\mu\leftrightarrow e$ operators.
Since 
the dipole   receives loop contributions from
almost all operators, it is sensitive to most
coefficients.  %In the $\mu\leftrightarrow e$ sector,
It also contributes to both  $\meee$ and $\mec$;
when the only contribution of a
coefficient to these processes is via the
dipole, 
 the sensitivity is in parentheses in tables
  \ref{tab:ls1} and  \ref{tab:qs},.

The $\mu\leftrightarrow e$  sensitivities are
given to three figures to mitigate
rounding uncertainty;  modulo  misprints and factors of 2,
they are expected to have  a $\sim 10\%$
uncertainty due to the one-loop QCD running, and 
a $\sim 50\%$  uncertainty for  scalar, tensor and
$GG$ operators in $\muc$, where the lattice and $\chi$PT determinations
of the scalar quark current in the nucleon differ by  $\sim 50\%$
\cite{Hoferichter:2015dsa,Lellouch}.

For comparaison, tables \ref{tab:ls1tau}, \ref{tab:qstau}
give estimated sensitivities of a selection of  LFV $\tau$ decays
to  operator coefficients  in the QED$\times$QCD-invariant basis.  In addition, table  \ref{tab:2l2qsSMEFTtau}
gives the sensitivities of some
 hadronic $\tau$ decays to $2\ell2q$ operators of  SMEFT
--- this illustrates, in the case
of  SMEFT operators involving quark doublets,
the cancellations  that can arise between the $u$ and $d$ quark
contributions in  $\tau$ decays to isotriplet mesons.

{\bf Note added:} During the completion of this manuscript,
appeared a comprehensive study of LFV in  hadronic $\tau$ decays
\cite{Husek}.  The authors calculate decays to a variety
of final states  using  $\chi$PT with resonances, and perform
some loop matching calculations  in  order to include
the interesting  operators ${\cal O}_{GG,Y}$ \cite{Celis:2014asa}. 
The  results in this manuscript are  less precise, include
$\tlg$ and $\tlll$ but fewer hadronic
decays, and include  QED loops in the RG running
up to the weak scale.  (QED effects are interesting, because
they can change the external legs and Lorentz
structure, transforming
difficult-to-detect operators into
well-constrained ones.)
Husek et al include the
numerically significant QCD running  via HEPfit
\cite{HEPFIT}, and obtain  constraints  an
SMEFT coefficients, and a correlation matrix.

\subsection{Including a selection of tau decays}
\label{app:tau}

A few  $\tau$ decays are included here, to
illustrate the  differences between
LFV  involving $\tau$s  and the $\mu\leftrightarrow e$
sector. Firstly, the experimental bounds are
more restrictive  for muon decays:
 $\widetilde{BR}(\mu\to e X) \lsim 10^{-12}$, to be
compared with $\widetilde{BR}(\tau\to \ell X) \lsim 10^{-7}$.
So there is sensitivity to smaller coefficients in
the muon sector. Furthermore, the loop contributions
of some operators involving heavy fermion ($\psi \in \{c,\tau,b\}$) bilinears,
can be enhanced by the  mass ratio $m_\psi/m_{\mu,\tau}$.
This, for instance, enhances the
sensitivity   to ${\cal O}^{cc}_{T,LL}$  of
$\meg$ % a better sensitivity than
with respect to $\teg$. %  to ${\cal O}^{cc}_{T,LL}$ operators.
Both these effects can be seen in the tables.

An advantage of  LFV  $\tau$ decays is 
%On the other hand
the  multitude of 
different hadronic final states, 
which  each constrain  a specific   combinations of quark  operator
coefficients.  This differs from $\mec$,
where   many  quark operators contribute
in interference. 
There are therefore  fewer ``flat directions''
for $\tau \to e$ than for  $\mu \to e$. This advantage is
exploited in \cite{Husek}, but not here, where only a few
decays are considered.
%This section outlines how the sensitivities to$\tau\leftrightarrow \ell$ operators, for $\ell \in \{e,\mu\}$, were estimated.
Also, only   $\tau\to e$ results
are listed; the sensitivities to $\tau \to \mu$
operators can be obtained by rescaling, as given in eqn (\ref{etomu}).

%so  the sensitivities of a few $\tau$ decays are included here to illustrate that

The limits here are a first attempt    to include QED loop effects
in  some LFV  $\tau $ decays, allowing to
estimate the sensitivity of these processes
to operators that  contribute via   log-enhanced loops.
%obtaining  an idea of the  ``one-at-a-time'' bounds on on  most  $\tau\leftrightarrow \ell$ coefficients at the weak scale.
Previous  works have included a  wider range of
$\tau$ decays,   but focus  mostly on 
operators that contribute at tree level.
%CITE MANY PEOPLE?
 The  prospects for discriminating 
among  operator
coefficients  by studying  asymmetries and angular
distributions  was studied in \cite{Kitano:2000fg,Matsuzaki:2007hh}
for leptonic decays (see also \cite{Giffels:2008ar,Dassinger:2007ru})
and using  hadronic decays was considered in
\cite{Celis:2014asa}.
Black etal \cite{BHHS} give 
the sensitivity of various rare decays to  a subset
of SMEFT coefficients that contribute at tree level;   a more complete
study  in  the SMEFT was performed recently in \cite{Husek}.
%Similar  tree-level ``one-at-a-time'' bounds have been tabulated elsewhere.
The intermediate-state contribution of  heavy quark  mesons
to  leptonic LFV  $\tau$ decays is considered in \cite{Kovalenko}.

Only  the $\Delta F=1$ decays such as $\teee$ are
included here; the operator basis and  anomalous dimension matrices
for such processes differ from the $\mu\to e$ case only
by permutation of lepton  flavour indices.
 $\Delta F=2$ decays (eg $\teme$)  would require
  additional operators and a dedicated RGE analysis.

Three  ingredients  are required  to calculate the
sensitivities tabulated here:
the experimental upper bounds on the BRs,
the theoretical  formulae for the  BRs as a function of a complete set
of operator coefficients at the experimental scale, and
the matrix which  accounts for loop contributions
by  expressing the  coefficients
at the experimental scale in terms of coefficients
at the weak scale. The experimental BRs
are in table \ref{tab:BRtau}, and the remainder of
this Appendix gives theoretical formulae for
the BRs. The matrix is  obtained by solving the
RGEs for the LFV operator coefficients (between $m_W$ and $m_\tau$),
and since we restrict to $\Delta F = 1$ operators,
these RGEs  are the same  as for 
 $\mu \to e$  decays and conversion\cite{PSI,meg@mW}
 (with some index changes).
A dedicated analysis would be required
to obtain reliable sensitivities, and ``observable-vectors''
for $\tau$-LFV.

 The $\tau$ can decay hadronically, so it is
convenient to rescale its LFV Branching Ratios :
\beq
\widetilde{BR}(\tau \to \ell X) \equiv
\frac{BR(\tau \to \ell X)}{BR(\tau \to \ell \bar{\nu} \nu)}~~~,
\label{tildeBRtau}
\eeq
where $BR(\tau \to \mu \bar{\nu} \nu) = .174$ and
 $BR(\tau \to e \bar{\nu} \nu) = .178$ \cite{PDB}.
The experimental bounds on the  considered  $\widetilde{BR}$s are given
 in table \ref{tab:BRtau}.

The masses of the final
state leptons are neglected in the rates. 
This simplification means that the  theoretical BRs
and RG evolution are identical for the $\tau \to \mu$ and
$\tau\to e$ sectors, after interchanging $\mu$
and $e$ indices.
Therefore  in the tables, only the four-fermion
operators describing $\tau\to e$ transitions are
listed; for operators with
two lepton indices, the sensitivities to $\tau \to \mu$  coefficients
can be obtained by rescaling:
\beq
C^{\mu\tau ...}_{Lor} \lsim C^{e\tau ...}_{Lor}
\sqrt{\frac{\widetilde{BR}(\tau \to \mu X)}{
\widetilde{BR}(\tau \to e X)}}
\label{etomu}
\eeq
where $...$ are the indices corresponding to the
final state $X$, and the Lorentz structure subscripts
should be identical for both coefficients. ICI
In the case of four-lepton operators,  all the $\mu$
and $e$ indices should be interchanged, {\it e.g.}
$C^{\mu\tau ee}_{Lor} \to C^{e\tau \mu\mu}_{Lor}$, and so on.

  \begin{table}
\renewcommand{\arraystretch}{1.25}
$\begin{array}{|c|c||c|c|}\hline 
\widetilde{BR} & \hbox{current ~ bound} &
\widetilde{BR} & \hbox{current ~ bound} \\
\hline
\tmg &  2.5 \times 10^{-7} %
&
\teg &  1.9 \times 10^{-7}%\cite{Babartmgteg}
\\
\tmmm &1.2 \times 10^{-7}%\cite{Hayasaka:2010np}
&\teee&  1.5 \times 10^{-7} %\cite{Hayasaka:2010np}
\\
\tau \to \mu \pi &6.3 \times 10^{-7}%\cite{Aubert:2006cz}
& \tau \to e \pi
&4.5 \times 10^{-7}%\cite{Miyazaki:2007jp}
\\
\tau \to \mu \eta &3.7 \times 10^{-7}%\cite{Miyazaki:2007jp}
& \tau \to e \eta
&5.2 \times 10^{-7}%\cite{Miyazaki:2007jp}
\\
\tau \to \mu \rho &6.9 \times 10^{-8}%\cite{Miyazaki:2011xe}
& \tau \to e \rho
&1.0 \times 10^{-7}%\cite{Miyazaki:2011xe}
\\
\hline
\end{array}$
\caption{Current bounds
  on selected $\tau$  lepton flavour violating branching ratios,
  from References \cite{Babartmgteg,Belletmg,Babartmgteg,Hayasaka:2010np,Aubert:2006cz,Miyazaki:2007jp,Miyazaki:2011xe}, normalised
to leptonic weak decays, as 
in eqn (\ref{tildeBRtau}). 
\label{tab:BRtau}}
\renewcommand{\arraystretch}{1.00}
\end{table}

The decays $\tmee$ and $\temm$ are mediated by
the $\tau\to e$ and $\tau \to \mu$ operators
considered here, but are not included
due to temporary discrepancies 
 in the  tensor contribution, between my calculation
and  \cite{Kitano:2000fg}.

 The  calculation of $\tau$  decays to mesons
is pedagogically introduced in \cite{BHHS}, and a
careful study considering many final states has
recently appeared \cite{Husek}. The decays
considered here are $\tau\to e \{\pi_0, \eta,\rho\}$;
the results for $\tau\to \mu \{\pi_0, \eta,\rho\}$
can be obtained as in eqn (\ref{etomu}).
These mesons are interesting because  they
probe complementary combinations of operator
coefficients at tree level.

The  decays to $\pi_0$ %and $ \eta$
mesons  probe
axial vector/pseudoscalar  operators, in % respectively
the isospin=1 combination $u -d$.
%and the combination $u$+$d$. 
In the
notation  of  \cite{PichLH}, where
$$
\langle 0|\overline{d} \g^\mu \g_5 u  |\pi^+(P) \rangle = iP^\mu \sqrt{2} f_\pi
~~~,~~~\Gamma(\tau \to \pi \nu) = \frac{G_F^2 f_\pi^2m_\tau^3}{8\pi}
$$
with  $f_\pi \simeq  92.2${ MeV}, the Branching Ratio in the presence of
(axial) vector operators is
\beq
\widetilde{BR}(\tau \to \ell \pi_0 ) =
\frac{3\pi^2f_\pi^2} { m_\tau^2}
|C^{uu}_{V,XR} - C^{uu}_{V,XL} -C^{dd}_{V,XR} + C^{dd}_{V,XL} |^2   ~~~,
\eeq
because 
\beq
\langle 0|J^\mu_{A-} |\pi^0(P) \rangle=
iP^\mu  f_\pi 
\eeq
where
$J^\mu_{A-}= \frac{1}{2}(\overline{u} \g^\mu \g_5 u
- \overline{d} \g^\mu \g_5 d)$, 
and  the coefficient of
$ 2 \sqrt{2} G_F (\bar{\ell}\g_\mu P_X\tau) J^\mu_{A-}$ in the
Lagrangian is\footnote{there is a factor 1/2  is missing in \cite{NSI3}.}
$C^{A-}=\frac{1}{2}(C^{uu}_{V,XR} - C^{uu}_{V,XL} -C^{dd}_{V,XR} + C^{dd}_{V,XL})$.
RG mixing vanishes for the axial current, but it
is renormalised (when attached to a chiral current) so
at $m_W$  this becomes the ``constraint''  
\bea
 2.3\times 10^{-3}&\gsim&
|C^{uu}_{V,XR} - C^{uu}_{V,XL} -C^{dd}_{V,XR} + C^{dd}_{V,XL} 
- \nonumber\\
&& \qquad
 \frac{\a_e}{\pi} (2C^{uu}_{V,XR} +2 C^{uu}_{V,XL} +C^{dd}_{V,XR} + C^{dd}_{V,XL} )
| \ln\frac{m_W}{m_\tau}
\label{pidec}
\eea
The correct  constraint, 
and ``observable-vector''(s) for this decay, 
could be obtained  from the expression for the BR in terms of all
coefficients that can contribute (including {\it e.g.}  the pseudoscalar
operators). However, eqn (\ref{pidec})  allows to calculate sensitivities,
and see  cancellations, such as  between
 $C^{uu}_{V,XL}$ and $ C^{dd}_{V,XL}$, due to which
 the SMEFT operators ${\cal O}_{EQ}$ and ${\cal O}_{LQ1}$
 do not contribute to $\tau \to \ell\pi_0$ at tree level.

It is interesting to also include  LFV $\tau$ decays
to the isosinglet $\eta$, because  there is
a contribution from $s$ quarks, and  not a cancellation
between the $u$s and $d$s. 
Still in the notation of \cite{PichLH},  with the %flavour-SU(3) 
%The $\eta$ is created by $\frac{1}{\sqrt{3}}(\overline{u} \g^\mu \g_5 u +\overline{d} \g^\mu \g_5 d - 2 \overline{s} \g^\mu \g_5 s)$
approximation  $f_\eta \sim f_\pi$  \cite{feta}, one obtains the
contribution
\beq
\widetilde{BR}(\tau \to \ell \eta ) =
\frac{ \pi^2f_\pi^2} { m_\tau^2}
|(C^{uu}_{V,XR} - C^{uu}_{V,XL}) +(C^{dd}_{V,XR} - C^{dd}_{V,XL}) - 2
(C^{ss}_{V,XR} - C^{ss}_{V,XL})|^2   ~~~.
\eeq
At $m_W$, this becomes
\bea
 4.4\times 10^{-3} &\gsim&
(C^{uu}_{V,XR} - C^{uu}_{V,XL}) +(C^{dd}_{V,XR} - C^{dd}_{V,XL}) - 2
(C^{ss}_{V,XR} - C^{ss}_{V,XL})\nonumber\\
&&
-\frac{\a_e}{\pi}(C^{dd}_{V,XR} + C^{dd}_{V,XL}
-2(C^{uu}_{V,XR} + C^{uu}_{V,XL}+ C^{ss}_{V,XR} + C^{ss}_{V,XL})
\ln\frac{m_W}{m_\tau}  \nonumber
\eea

Pseudoscalar operators can also contribute to  the decays
$\tau \to \ell \pi_0, \ell \eta$.  The operator expectation
values
 \bea
  \langle 0|\frac{1}{2}(\bar{u}\g_5u - \bar{d}\g_5d) |\pi_0\rangle &=& \frac{f m_\pi^2}{(m_u + m_d)} \nonumber\\
  \langle 0|\frac{1}{2\sqrt{3}}(\bar{u}\g_5u + \bar{d}\g_5d - 2\bar{s}\g_5s )|\eta\rangle& =& \frac{f m_\pi^2}{(m_u + m_d)} = \frac{3 f_\eta m_\eta^2}{(m_u + m_d+ 4 m_s)} \label{pseudoscalarchi}
  \eea
give a contribution of pseudoscalar coefficients to the  Branching Ratios of
\bea
\widetilde{BR} (\tau \to e\{\pi_0,\eta\}) =
96\pi^2 \left(\frac{m_{\pi_0}}{m_\tau}\right)^4
\left(\frac{f_\pi}{m_u+m_d}\right)^2 |C_{\eta,\pi}|^2
%~~~,~~~({\rm pseudoscalar ~operators})
%\left(\frac{}{}\right)
\label{where}
\eea
where  in the normalisation  of eqn  (\ref{L}),
the coefficients of the operators of
eqn (\ref{pseudoscalarchi}) are
 $C_{\pi} = \frac{1}{2}( C^{uu}_{S,XR} - C^{uu}_{S,XL}  - C^{dd}_{S,XR} + C^{dd}_{S,XL})$, and
   $C_{\eta}=\frac{1}{2\sqrt{3}}( C^{uu}_{S,XR} - C^{uu}_{S,XL}
   +C^{dd}_{S,XR} - C^{dd}_{S,XL} - 2 C^{ss}_{S,XR} + C^{ss}_{S,XL})$.
QED loops can mix tensor operators into (pseudo)scalars, so
this will give some sensitivity to the $u, d, s$ tensor
operators. 

Finally, decays to the vector $\rho$ meson are normalised to
$BR(\tau\to \rho\nu)$, assuming  $\rho \to \pi\pi$ (as in \cite{NSI3}; see
\cite{Husek} for a more sophisticated solution)
and   with the usual factor of $2$ for
the normalisation of neutral and charged particles:
\bea
\widetilde{BR}(\tau \to \ell \rho_0 ) &\approx&
\frac{BR(\tau \to \nu \rho)}
{BR(\tau \to \ell \nu \bar{\nu})}
\frac{\Gamma(\tau \to \ell \rho_0)}
{\Gamma(\tau \to \nu \rho)}  \nonumber\\
&\approx& 1.43 \frac{|(C^{uu}_{V,XR} + C^{uu}_{V,XL}) -(C^{dd}_{V,XR} + C^{dd}_{V,XL})|^2}
{8 |V_{ud}|^2}
\eea
In the second expression,  the contribution
of the dipole operator (analogous to the dipole
contribution to $\teee$) is neglected, because the
current experimental bounds on $\teg$ and
$\tau \to e \rho_0$ are comparable. 
QED loops mix    vector operators  of different
quark flavour via penguin diagrams, giving  this  decay
 some sensitivity to  to the coefficients at $m_W$
 of vector operators with a
heavy quark current:
\bea
7.5\times 10^{-4} &\gsim & (C^{uu}_{V,XR} + C^{uu}_{V,XL}) -(C^{dd}_{V,XR} + C^{dd}_{V,XL})\nonumber\\ &&
+ \frac{\a_e}{3\pi} ( 2C^{uu}_{V,XL} -10 C^{uu}_{V,XR}
-( C^{dd}_{V,XR} -5C^{dd}_{V,XL})
  )\ln \frac{m_W}{m_\tau} 
\nonumber\\ && +
  \frac{2\a_e}{3\pi} ( 
     2 C^{ee}_{V,XL} +  C^{ee}_{V,XR} +
 2 C^{\tau\tau}_{V,XL} +  C^{\tau\tau}_{V,XR})\ln
  \frac{  m_W}{m_\tau} 
     \label{bbdrho}\\
&&
+ \frac{2\a_e}{3\pi}(C^{\mu\mu}_{V,XR} + C^{\mu\mu}_{V,XL}  + C^{ss}_{V,XR} + C^{ss}_{V,XL}
- 2(C^{cc}_{V,XR} + C^{cc}_{V,XL})) \ln
  \frac{  m_W}{m_\tau}
\nonumber\\&&
+ \frac{2\a_e}{3\pi}(C^{bb}_{V,XR} + C^{bb}_{V,XL})
\nonumber
\eea

\begin{table}[ht]
\begin{center}
\begin{tabular}{|l|l|l|l|}
\hline
coefficient & $\meg$ & $\meee$ &$\muc$ \\
 \hline
$|C_{D,X}|$ &$1.12 \times 10^{-8}$ &  $2.21 \times 10^{-7}$  & $2.35 \times 10^{-7}$   \\
$|C_{GG,X}|$ &    &      & $5.3 \times 10^{-7}$   \\
$|C^{ee}_{V,XX}|$&$1.10 \times 10^{-4}$
  &$7.80\times 10^{-7}$
    & $1.86\times 10^{-5}$ \\
$|C^{ee}_{V,XY}|$ &$ 2.55\times 10^{-4}$
  & $9.31\times 10^{-7}$
    &$3.77\times 10^{-5}$\\
$|C^{ee}_{S,XX}|$& $1.73\times 10^{-4}$
  &$2.8\times 10^{-6}$
   &  ($3.64\times 10^{-3}$)\\&   &&\\
$|C^{\m\m}_{V,XX}|$& $1.10\times 10^{-4}$ 
  &$5.64\times 10^{-5}$
   & $1.85\times 10^{-5}$ \\
$|C^{\m\m}_{V,XY}|$& $2.56\times 10^{-4}$
  &$1.11\times 10^{-4}$
   & $3.77\times 10^{-5}$ \\
$|C^{\m\m}_{S,XX}|$& $8.24\times 10^{-7}$
  &($1.63\times 10^{-5}$)
   & ($1.73\times 10^{-5}$)  \\
      &  &&\\
$|C^{\tau\tau}_{V,XX}|$&  $3.84\times 10^{-4}$
  &$1.94\times 10^{-4}$
   & $3.72\times 10^{-5}$ \\
$|C^{\tau\tau}_{V,XY}|$&$4.45\times 10^{-4}$
  &$1.94\times 10^{-4}$
   & $3.72\times 10^{-5}$ \\
$|C^{\tau\tau}_{S,XX}|$& $5.33\times 10^{-6}$
  & $(1.05\times 10^{-4})$
   &  $(1.12\times 10^{-4})$ \\
$|C^{\tau\tau}_{S,XY}|$&  $3.62\times 10^{-5}$
  &   $(7.28\times 10^{-4})$
   &   $(7.75\times 10^{-4})$  \\
$|C^{\tau\tau}_{T,XX}|$& $1.07\times 10^{-8}$
  &($2.11\times 10^{-7}$)
   & ($2.25\times 10^{-7}$)  \\% &   &&\\
\hline
\hline
\end{tabular}
\caption{Current sensitivities of
the processes in the first row to  the
 coefficients, at $m_W$,
 of QCD$\times$QED-invariant 2- and 4-lepton operators
 defined in section \ref{ssec:ops}.
 $X,Y \in \{L,R\}, X\neq Y$.
\label{tab:ls1}}
\end{center}
\end{table}

\begin{table}[ht]
\begin{center}
\begin{tabular}{|l|l|l|l|}
\hline
coefficient & $\meg$ & $\meee$ &$\muc$ \\
 \hline
$|C^{dd}_{V,XX}|$& $1.04\times 10^{-3}$ 
  &$2.03\times 10^{-4}$
  & $5.40\times 10^{-8}$ \\
$|C^{dd}_{V,XY}|$&$1.64\times 10^{-3}$
  &$2.01\times 10^{-4}$
   & $5.30\times 10^{-8}$ \\
$|C^{dd}_{S,XX}|$&$5.79\times 10^{-3}$
  &$(1.14\times 10^{-1})$
   & $1.03\times 10^{-8}$ \\
$|C^{dd}_{S,XY}|$& 
   &
   & $1.03\times 10^{-8}$ \\    
$|C^{dd}_{T,XX}|$&$5.57\times 10^{-6}$
  &($1.10\times 10^{-4}$)
   & $1.90\times 10^{-7}$ \\    
&&&\\
$|C^{uu}_{V,XX}|$&$3.59\times 10^{-4}$
  &$1.00\times 10^{-4}$
   & $6.03\times 10^{-8}$ \\
$|C^{uu}_{V,XY}|$& $2.87\times 10^{-4}$
   &$1.02\times 10^{-4}$
    & $6.25\times 10^{-8}$ \\
$|C^{uu}_{S,XX}|$&$3.09\times 10^{-3}$
  &$(6.71\times 10^{-1})$
   & $1.03\times 10^{-8}$ \\
$|C^{uu}_{S,XY}|$& 
   &
    & $1.03\times 10^{-8}$ \\    
$|C^{uu}_{T,XX}|$&$5.95\times 10^{-6}$
  &($1.17\times 10^{-4}$)
   & $9.65\times 10^{-8}$ \\    
&&&\\
 $|C^{ss}_{V,XX}|$& $1.01\times 10^{-3}$
  &$2.03\times 10^{-4}$
    & $3.73\times 10^{-5}$ \\
$|C^{ss}_{V,XY}|$& $1.64\times 10^{-3}$
   &$2.01\times 10^{-4}$
     & $3.73\times 10^{-5}$ \\
$|C^{ss}_{S,XX}|$&$2.92\times 10^{-4}$
  &$(5.77\times 10^{-3})$
   & $2.13\times 10^{-7}$ \\
$|C^{ss}_{S,XY}|$&
$1.41\times 10^{-2}$
  &$(2.78\times 10^{-1})$
    & $2.13\times 10^{-7}$ \\    
$|C^{ss}_{T,XX}|$&$2.82\times 10^{-7}$
  &($5.56\times 10^{-6}$)
   & $2.33\times 10^{-6}$ \\    
&&&\\   
$|C^{cc}_{V,XX}|$& $3.59\times 10^{-4}$
  &$8.99\times 10^{-5}$
   & $1.68\times 10^{-5}$ \\
$|C^{cc}_{V,XY}|$& $2.87\times 10^{-4}$
  &$9.05\times 10^{-5}$
    &  $1.67\times 10^{-5}$  \\
$|C^{cc}_{S,XX}|$&$5.23\times 10^{-6}$
  &$(1.03\times 10^{-4})$
   & $1.83\times 10^{-6}$ \\
$|C^{cc}_{S,XY}|$& 
   $2.37\times 10^{-5}$
  &$(4.68\times 10^{-4})$
    & $1.80\times 10^{-6}$ \\    
$|C^{cc}_{T,XX}|$&$1.01\times 10^{-8}$
  &($1.99\times 10^{-7}$)
   & $2.13\times 10^{-7}$ \\    
&&&\\   
$|C^{bb}_{V,XX}|$& $1.32\times 10^{-3}$
  &$2.56\times 10^{-4}$
    & $4.71\times 10^{-5}$ \\
$|C^{bb}_{V,XY}|$&$2.05\times 10^{-3}$
   &$2.54\times 10^{-4}$
     & $4.71\times 10^{-5}$ \\
   $|C^{bb}_{S,XX}|$&$1.01\times 10^{-5}$
  &$(1.98\times 10^{-4})$
   & $7.10\times 10^{-6}$ \\
$|C^{bb}_{S,XY}|$&
$4.04\times 10^{-5}$
  &$(7.98\times 10^{-4})$
    & $6.92\times 10^{-6}$ \\    
$|C^{bb}_{T,XX}|$&$7.81\times 10^{-9}$
  &($1.52\times 10^{-7}$)
   & $1.64\times 10^{-7}$ \\      
\hline
\hline
\end{tabular}
\caption{
Current sensitivities of
the processes in the first row to  the
 coefficients, evaluated at $m_W$,
 of QCD$\times$QED-invariant 2-lepton-2quark operators
 defined in section \ref{ssec:ops}.
 $X,Y \in \{L,R\}, X\neq Y$.
\label{tab:qs}}
\end{center}
\end{table}

%SMEFTTABLES
\begin{table}[ht]
\begin{center}
\begin{tabular}{|l|l|l|l|}
\hline
coefficient & $\meg$ & $\meee$ &$\muc$ \\
 \hline
$C^{e\mu}_{e\g},C^{\mu e~*}_{e\g}$ &$1.12 \times 10^{-8}$ &  $2.21 \times 10^{-7}$  & $2.35 \times 10^{-7}$   \\
$C^{e\mu}_{HE}$ &$1.18 \times 10^{-5}$
  &$1.20\times 10^{-6}$
   & $1.42\times 10^{-7}$ \\
$C^{e\mu}_{HL1}$ &$ 1.57\times 10^{-5}$
  &$1.17\times 10^{-6}$
   & $1.61\times 10^{-7}$ \\
 $C^{e\mu}_{HL3}$ &$1.57 \times 10^{-5}$
  &$1.17\times 10^{-6}$
   & $1.61\times 10^{-7}$ \\  
$C^{e\mu }_{EH}$ &$ 7.52\times 10^{-7}$
  &$1.48\times 10^{-5}$
   & $5.36\times 10^{-5}$ \\
%   & $4.89\times 10^{-5}$ \\ agrees DKUY?   
\hline
\hline
\end{tabular}
\caption{Current sensitivities of  the
processes in the first row to  the
 coefficients of SMEFT operators
 at $m_W$, added to the
 Lagrangian as in eqn(\ref{SMEFT}).
 $C^{e\mu}_{e\g} = c_WC^{e\mu}_{EB}-s_WC^{e\mu}_{EW}$.
\label{tab:2lsSMEFT}}
\end{center}
\end{table}

\begin{table}[ht]
\begin{center}
\begin{tabular}{|l|l|l|l|}
\hline
coefficient & $\meg$ & $\meee$ &$\muc$ \\
 \hline
$\Sigma C^{e\mu ee}_{EE}, \Sigma C^{e\mu ee}_{LL}$ &$ 1.10\times 10^{-4}$
  &$7.87\times 10^{-7}$
   & $1.85\times 10^{-5}$ \\
$C^{eee\mu}_{LE}$,$C^{e\mu ee}_{LE}$ &$2.55 \times 10^{-4}$
   & $9.31\times 10^{-7}$
   &$3.77\times 10^{-5}$
  \\
$\Sigma C^{e\mu \mu \mu}_{EE},\Sigma C^{e\mu \mu \mu}_{LL}$
&$1.10 \times 10^{-4}$
  &$5.67\times 10^{-5}$
   & $1.85\times 10^{-5}$ \\
 $C^{\mu \mu e\mu}_{LE},C^{e\mu \mu \mu}_{LE}$
 &$2.55 \times 10^{-4}$
  &$1.11\times 10^{-4}$
   & $3.77\times 10^{-5}$ \\
   $\Sigma C^{e\mu \tau\tau}_{EE},\Sigma C^{e\mu \tau\tau}_{LL}$
   &$3.84 \times 10^{-4}$
  &$1.95\times 10^{-4}$
   & $3.72\times 10^{-5}$ \\
   $\Sigma C^{e  \tau\tau \mu}_{LL}$
   &$3.84 \times 10^{-4}$
  &$1.95\times 10^{-4}$
   & $3.72\times 10^{-5}$ \\
$C^{\tau\tau e\mu}_{LE},C^{e\mu \tau\tau}_{LE}$
&$4.40 \times 10^{-4}$
  &$1.91\times 10^{-4}$
   & $3.75\times 10^{-5}$ \\
   $C^{e\tau\tau \mu}_{LE},C^{ \tau \mu e\tau}_{LE}$
&$1.80 \times 10^{-5}$
  &$3.64\times 10^{-4}$
   & $3.88\times 10^{-4}$ \\
 \hline
\hline
\end{tabular}
\caption{
Current sensitivities of  the
processes in the first row to  the
 coefficients of SMEFT operators
 at $m_W$, added to the
 Lagrangian as in eqn(\ref{SMEFT}).
  $\sum C_{LL}^{\zeta}$ and
  $\sum C_{RR}^{\zeta}$ are defined in eqn (\ref{eqsumRR}).
 %$C^{e\mu}_{e\g} = c_WC^{e\mu}_{EB}-s_WC^{e\mu}_{EW}$.
\label{tab:4lsSMEFT}}
\end{center}
\end{table}

\begin{table}[ht]
\begin{center}
\begin{tabular}{|l|l|l|l|}
\hline
coefficient & $\meg$ & $\meee$ &$\muc$ \\
 \hline
%u,d
$C^{e\mu uu}_{EU}$ &$3.59 \times 10^{-4}$
  &$1.00\times 10^{-4}$
   & $6.03\times 10^{-8}$ \\
    $C^{e\mu uu}_{LU}$ &$2.87 \times 10^{-4}$
  &$1.02\times 10^{-4}$
   & $6.25\times 10^{-8}$ \\
      $C^{e\mu uu}_{LEQU}$ &$3.09 \times 10^{-3}$
  &$6.71\times 10^{-1}$
   & $1.03\times 10^{-8}$ \\
         $C^{e\mu uu}_{T,LEQU}$ &$5.95 \times 10^{-6}$
  &$1.17\times 10^{-4}$
   & $9.65\times 10^{-8}$ \\
%% &&&\\%d 
$C^{e\mu dd}_{LQ1}$ &$2.67 \times 10^{-4}$
  &$1.98\times 10^{-4}$
   & $2.85\times 10^{-8}$ \\
      $C^{e\mu dd}_{LQ3}$ &$5.47 \times 10^{-4}$
&$6.71 \times 10^{-5}$
   &  $5.09\times 10^{-7}$\\
    $C^{e\mu dd}_{EQ}$ &$2.44\times 10^{-4}$
  &$2.05\times 10^{-4}$
   & $2.87\times 10^{-8}$ \\
$C^{e\mu dd}_{ED}$ &$1.04 \times 10^{-3}$
  &$2.03\times 10^{-4}$
   & $5.40\times 10^{-8}$ \\
      $C^{e\mu dd}_{LD}$
      &$1.64 \times 10^{-3}$
  &$2.01\times 10^{-4}$
   &  $5.30\times 10^{-8}$ \\
        $C^{e\mu dd}_{LEDQ}$ &
  &
   & $1.01\times 10^{-8}$ \\
&&&\\
%c,s
$C^{e\mu cc}_{EU}$ &$3.59 \times 10^{-4}$
  &$8.99\times 10^{-5}$
   & $1.68\times 10^{-5}$\\
    $C^{e\mu cc}_{LU}$ &$ 2.87\times 10^{-4}$
  &$9.10\times 10^{-5}$
   & $1.67\times 10^{-5}$ \\
      $C^{e\mu cc}_{LEQU}$ &$ 5.23\times 10^{-5}$
  &$1.03\times 10^{-4}$
   & $1.83\times 10^{-6}$ \\
         $C^{e\mu cc}_{T,LEQU}$ &$1.01 \times 10^{-8}$
  &$1.99\times 10^{-7}$
   & $2.09\times 10^{-7}$ \\
 %
% &&&\\
 $C^{e\mu ss}_{LQ1}$ &$2.67 \times 10^{-4}$
  &$1.60\times 10^{-4}$
   & $3.02\times 10^{-5}$\\
   $C^{e\mu ss}_{LQ3}$ &$5.47 \times 10^{-4}$
  &$6.20\times 10^{-5}$
   &  $1.15\times 10^{-5}$\\
   $C^{e\mu ss}_{EQ}$ &$ 2.44\times 10^{-4}$
  &$1.64\times 10^{-4}$
   & $2.98\times 10^{-5}$ \\
%s
$C^{e\mu ss}_{ED}$ &$ 1.04\times 10^{-3}$
  &$2.03\times 10^{-4}$
   & $3.73\times 10^{-5}$ \\
       $C^{e\mu ss}_{LD}$ &
      $1.64 \times 10^{-3}$
  &$2.01\times 10^{-4}$
   &  $3.73\times 10^{-5}$\\
        $C^{e\mu ss}_{LEDQ}$ &
        $1.41\times 10^{-2}$
  &$2.78\times 10^{-2}$
   & $2.09\times 10^{-7}$ \\   
&&&\\
%b
$C^{e\mu bb}_{ED},C^{e\mu bb}_{LQ1}$
&$1.32\times 10^{-3}$
  &$2.56\times 10^{-4}$
   & $4.71\times 10^{-5}$ \\
   $C^{e\mu bb}_{LQ3}$ &$1.32 \times 10^{-3}$
  &$2.56\times 10^{-4}$
   & $4.71\times 10^{-5}$ \\
      $C^{e\mu bb}_{LD},C^{e\mu bb}_{EQ}$
      &$2.07 \times 10^{-3}$
  &$2.51\times 10^{-4}$
   & $4.72\times 10^{-5}$ \\
        $C^{e\mu bb}_{LEDQ}$ &
        $4.04\times 10^{-5}$
  &$7.98\times 10^{-4}$
   & $6.92\times 10^{-6}$ \\  
\hline
\hline
\end{tabular}
\caption{Current sensitivities of  $\meg$,
$\meee$,  and $\muc$  to  the
 coefficients of SMEFT operators
 at $m_W$.
\label{tab:2l2qsSMEFT}}
\end{center}
\end{table}

\begin{table}[ht]
\begin{center}
\begin{tabular}{|l|l|l|l|}
\hline
coefficient & $\teg$ & $\teee$ &$\tau \to e \rho$ \\
 \hline
$|C_{D,X}|$ &$7.35 \times 10^{-6}$ &  $6.36 \times 10^{-5}$  & $...$   \\
%$|C_{GG,X}|$ & ---   &  ---    & $5.3 \times 10^{-7}$   \\
$|C^{ee}_{V,XX}|$&$1.29\times 10^{-1}$
  &$2.83 \times 10^{-4}$
    & $6.55\times 10^{-2}$ \\
$|C^{ee}_{V,XY}|$ &$ 3.00\times 10^{-1}$
  & $3.78\times 10^{-4}$
    &$1.31 \times 10^{-1}$\\
$|C^{ee}_{S,XX}|$& $3.37$
  &$(1.06\times 10^{-3})$
   & \\&   &&\\
$|C^{\m\m}_{V,XX}|$& $2.59\times 10^{-1}$ 
  &$7.65\times 10^{-2}$
   & $1.31\times 10^{-1} $ \\
$|C^{\m\m}_{V,XY}|$& $3.00\times 10^{-1}$
  &$7.49\times 10^{-2}$
   & $1.31\times 10^{-1}$ \\
$|C^{\m\m}_{S,XX}|$& $9.76\times 10^{-1}$
  &($8.46$)
   & \\ %($\times 10^{-5}$)  \\
 $|C^{\m\m}_{S,XY}|$& $4.17\times 10^{-1}$
  &($3.61$)
   & \\%($\times 10^{-5}$)  \\
   $|C^{\mu\mu}_{T,XX}|$& $2.03\times 10^{-3}$
  &($1.76\times 10^{-2}$)
   & \\%($\times 10^{-}$)  \\% &   &&\\
      &  &&\\
$|C^{\tau\tau}_{V,XX}|$&  $1.29\times 10^{-1}$
  &$3.82\times 10^{-2}$
   & $6.55\times 10^{-2}$ \\
$|C^{\tau\tau}_{V,XY}|$&$3.00\times 10^{-1}$
  &$7.49\times 10^{-2}$
   & $1.31\times 10^{-1}$ \\
$|C^{\tau\tau}_{S,XX}|$& $9.68\times 10^{-4}$
& $(8.39\times 10^{-3})$
   & \\% $\times 10^{-4}$ \\
%$|C^{\tau\tau}_{S,XY}|$&  $3.62\times 10^{-5}$  &   $(7.28\times 10^{-4})$   &   $(7.75\times 10^{-4})$  \\
%
\hline
\hline
\end{tabular}
\caption{
Current sensitivities of
the processes in the first row, to  the
 coefficients $C^{e\tau ...}_{Lor}$, evaluated at $m_W$,
 of $\tau \leftrightarrow e$ flavour-changing,
 QCD$\times$QED-invariant 2- and 4-lepton operators,
 defined as in section \ref{ssec:ops} with
 $\mu \to \tau$.
 $X,Y \in \{L,R\}, X\neq Y$.
\label{tab:ls1tau}}
\end{center}
\end{table}

\begin{table}[ht]
\begin{center}
\begin{tabular}{|l|l|l|l|l|l|}
\hline
coefficient & $\teg$ & $\teee$ &$\tau \to e \rho$&$\tau \to e \pi$& $\tau \to e \eta$\\
 \hline
$|C^{dd}_{V,XX}|$& $7.03\times 10^{-1}$ 
  &$7.84\times 10^{-2}$
  & $7.61\times 10^{-4}$%(h=\rho)$
  & $2.32\times 10^{-3}$& $4.44\times 10^{-3}$\\
$|C^{dd}_{V,XY}|$&$1.10$
  &$7.79\times 10^{-2}$
   & $7.48\times 10^{-4}$
    & $2.28\times 10^{-3}$& $4.36\times 10^{-3}$\\
$|C^{dd}_{S,XX}|$&$66.0$
  &$(572)$
   &  &$3.74\times 10^{-4}$&$6.96\times 10^{-4}$ \\
$|C^{dd}_{S,XY}|$& 
   &
   &  & $3.74\times 10^{-4}$&$6.96\times 10^{-4}$ \\    
$|C^{dd}_{T,XX}|$&$6.28\times 10^{-2}$
  &($5.44\times 10^{-1}$)
   &  &$6.92\times 10^{-3}$&$1.29\times 10^{-2}$ \\
&&&&&\\
$|C^{uu}_{V,XX}|$&$2.42\times 10^{-1}$
  &$3.88\times 10^{-2}$
   & $7.46\times 10^{-4}$
   &$2.26\times 10^{-3}$ &$4.33\times 10^{-3}$\\
   %(h=\rho)$ \\
$|C^{uu}_{V,XY}|$& $1.94\times 10^{-1}$
   &$3.93\times 10^{-2}$
    & $7.72\times 10^{-4}$% (h=\rho)$ \\
&$2.34\times 10^{-3}$ &$4.48\times 10^{-3}$\\
$|C^{uu}_{S,XX}|$&$35.3$
  &$(305)$&
    &$3.73\times 10^{-4}$&$6.93\times 10^{-4}$ \\
$|C^{uu}_{S,XY}|$&    &
  & &$3.73\times 10^{-4}$&$6.93\times 10^{-4}$  \\    
$|C^{uu}_{T,XX}|$&$6.71\times 10^{-2}$
  &($5.81\times 10^{-1}$)
   &  &$3.46\times 10^{-3}$&$6.44\times 10^{-3}$ \\ 
&&&&&\\
$|C^{ss}_{V,XX}|$& $7.03\times 10^{-1}$
  &$7.84\times 10^{-2}$
    & $1.31\times 10^{-1}$
  &  &$2.22\times 10^{-3}$\\
$|C^{ss}_{V,XY}|$& $1.11$
   &$7.79\times 10^{-2}$
     & $1.31\times 10^{-1}$ 
   &  &$2.18\times 10^{-3}$\\
$|C^{ss}_{S,XX}|$&$3.23$
  &$(28.0)$
   &&&$3.48\times 10^{-4}$ \\
$|C^{ss}_{S,XY}|$&
$15.9$
  &$(138)$
    &  &&$3.48\times 10^{-4}$ \\    
$|C^{ss}_{T,XX}|$&$3.08\times 10^{-3}$
  &($2.66\times 10^{-2}$)
   && &$6.44\times 10^{-3}$ \\      
&&&&&\\    
$|C^{cc}_{V,XX}|$& $2.42\times 10^{-1}$
  &$3.46\times 10^{-2}$
   & $5.84\times 10^{-2}$&&\\ % (h=\rho)$ \\
$|C^{cc}_{V,XY}|$& $1.94\times 10^{-1}$
  &$3.50\times 10^{-2}$
    &  $5.84\times 10^{-2} $&&  \\
$|C^{cc}_{S,XX}|$&$6.11\times 10^{-2}$
  &$(5.29\times 10^{-1})$
   & & &\\
$|C^{cc}_{S,XY}|$& 
   $2.68\times 10^{-1}$
  &($2.32$)
    & && \\    
$|C^{cc}_{T,XX}|$&$1.06\times 10^{-4}$ 
  &($1.01\times 10^{-3}$)
   & && \\    
&&&&&\\    
$|C^{bb}_{V,XX}|$& $8.80\times 10^{-1}$
  &$9.80\times 10^{-2}$
    & $1.64\times 10^{-1}$& &\\
$|C^{bb}_{V,XY}|$&$1.38$
   &$9.73\times 10^{-2}$
     & $1.64\times 10^{-1}$& &\\
   $|C^{bb}_{S,XX}|$&$(1.16\times 10^{-1})$
  &$(1.00)$
   & &&\\
$|C^{bb}_{S,XY}|$&
$4.57\times 10^{-1}$
  &$3.96$
    & &&\\    
$|C^{bb}_{T,XX}|$&$8.82\times 10^{-5}$
  &($7.65$)
   & & &\\      
\hline
\hline
\end{tabular}
\caption{
Similar to table \ref{tab:ls1tau},
but for  2-lepton-2-quark operators.
%Current sensitivities of  $\teg$,$\teee$,  and $\tau \to \{\rho,\pi,\eta\}$  to  thecoefficients   evaluated at  $m_W$, of   $\tau \leftrightarrow e$QCD$\times$QED-invariant operators.  $X,Y \in \{L,R\}, X\neq Y$.
\label{tab:qstau}}
\end{center}
\end{table}

\begin{table}[ht]
\begin{center}
\begin{tabular}{|l|l|l|l|l|}
\hline
coefficient &$\teg$& $\tau \to e \rho$ &  $\tau \to e \pi$ & $\tau \to e \eta$ \\
 \hline
%u 
$C^{e\tau dd}_{LQ1}$ &$1.80 \times 10^{-1}$ &$3.74 \times 10^{-2}$
  &$8.93\times 10^{-2}$
   & $2.19\times 10^{-3}$ \\%
   $C^{e\tau dd}_{LQ3}$  &$3.69 \times 10^{-1}$ &$3.77 \times 10^{-4}$
  &$1.14\times 10^{-3}$
   &  $1.71\times 10^{-1}$\\%
$C^{e\tau dd}_{LD}$ &$1.10$ &$7.48 \times 10^{-4}$
  &$2.28\times 10^{-3}$
   &  $4.36\times 10^{-3}$ \\%
$   C^{e\tau dd}_{EQ}$ &$1.65 \times 10^{-1}$ &$2.38 \times 10^{-2}$
  &$8.93\times 10^{-2}$
   & $2.21\times 10^{-3}$ \\
 $C^{e\tau dd}_{ED}$ &$7.04 \times 10^{-1}$ &$7.61\times 10^{-4}$
  &$2.32\times 10^{-3}$
   &  $4.44 \times 10^{-3}$\\%
   $C^{e\tau dd}_{LEDQ}$ &&
  &$3.74\times 10^{-4}$
   & $6.96\times 10^{-4}$ \\ 
$C^{e\tau uu}_{EU}$ &$2.42 \times 10^{-1}$ &$7.46 \times 10^{-4}$
  &$2.26\times 10^{-3}$
   & $4.33\times 10^{-3}$ \\
$C^{e\tau uu}_{LU}$ &$1.94 \times 10^{-1}$ &$7.72 \times 10^{-4}$
  &$2.34\times 10^{-3}$
   & $4.48\times 10^{-3}$ \\
$C^{e\tau uu}_{LEQU}$ &35.3&
  &$3.73\times 10^{-4}$
   & $6.93\times 10^{-4}$ \\
$C^{e\tau uu}_{T,LEQU}$  &$6.71 \times 10^{-2}$&
  &$3.46\times 10^{-3}$
   & $6.44\times 10^{-3}$ \\
&&&&\\ %s,c
$C^{e\tau ss}_{LQ1}$ &$1.80 \times 10^{-1}$ &$1.05 \times 10^{-1}$
  &&$2.22 \times 10^{-3}$ \\
   $C^{e\tau ss}_{LQ3}$ &$3.69 \times 10^{-1}$ &$4.04 \times 10^{-2}$
  &&$2.22 \times 10^{-3}$\\
$C^{e\tau ss}_{LD}$ &1.06&
      $1.31 \times 10^{-1}$
  &&$2.18 \times 10^{-3}$\\
   $C^{e\tau ss}_{EQ}$ &$1.65 \times 10^{-1}$ &$ 1.05\times 10^{-1}$
  &   &$2.18 \times 10^{-3}$\\
$C^{e\tau ss}_{ED}$ &$7.04 \times 10^{-1}$  &$ 1.31\times 10^{-1}$
  &&$2.22 \times 10^{-3}$\\
           $C^{e\tau ss}_{LEDQ}$&16 &&&$3.48\times 10^{-4}$\\
$C^{e\tau cc}_{EU}$ &$2.42 \times 10^{-1}$ &$5.84 \times 10^{-2}$
  &&\\
$C^{e\tau cc}_{LU}$ &$1.94 \times 10^{-1}$ &$ 5.84\times 10^{-2}$
  &   &\\
      $C^{e\tau cc}_{LEQU}$  &$6.11 \times 10^{-2}$&
&&\\      
         $C^{e\tau cc}_{T,LEQU}$ &$1.16 \times 10^{-4}$ &&&\\
&&&&\\ %b
$C^{e\tau bb}_{LQ1}$ &$8.79 \times 10^{-1}$
&$1.64\times 10^{-1}$
 &&\\
   $C^{e\tau bb}_{LQ3}$ &$8.79 \times 10^{-1}$ &$1.32 \times 10^{-3}$
 &&\\
$C^{e\tau bb}_{LD}$&1.38
      &$1.64 \times 10^{-1}$  &&\\
$C^{e\tau bb}_{EQ}$ &$8.79 \times 10^{-1}$
      &$1.64 \times 10^{-1}$  &&\\
   $C^{e\tau bb}_{ED}$
 &$8.79 \times 10^{-1}$&$1.64\times 10^{-1}$
 &&\\           
        $C^{e\tau bb}_{LEDQ}$ &$4.57 \times 10^{-1}$ &  &&\\%
\hline
\hline
\end{tabular}
\caption{Current sensitivities of selected  hadronic
$\tau$ decays to  the
 coefficients evaluated at $m_W$  of $2\ell2q$  SMEFT operators
\label{tab:2l2qsSMEFTtau}}
\end{center}
\end{table}

\end{document}